\documentclass[a4paper]{llncs}

\title{Solving Parity Games on Integer Vectors
\thanks{
Technical Report EDI-INF-RR-1417 of the School of Informatics at the University
of Edinburgh, UK. (http://www.inf.ed.ac.uk/publications/report/).
Full version (including proofs) of material presented at CONCUR 2013 (Buenos Aires, Argentina). arXiv.org - CC BY 3.0.
}
}

\author{Parosh Aziz Abdulla\inst{1}\thanks{
Supported by Uppsala Programming for Multicore
Architectures Research Center (UpMarc).
}
\and Richard Mayr\inst{2}\thanks{
Supported by Royal Society grant IE110996.}
\and Arnaud Sangnier\inst{3}
\and Jeremy Sproston\inst{4}\thanks{
Supported by the project AMALFI (University of Turin/Compagnia di San Paolo) and the MIUR-PRIN project CINA.}}
\institute{Uppsala University, Sweden
\and University of Edinburgh, UK
\and LIAFA, Univ Paris Diderot, Sorbonne Paris Cit\'e, CNRS, France
\and University of Turin, Italy}

\date{}

\usepackage{pslatex}

\usepackage{graphics}
\usepackage{wrapfig}
\usepackage{amscd}
\usepackage{enumerate}
\usepackage{latexsym}
\usepackage{amssymb,amsmath}
\usepackage{xspace}

\usepackage[ruled,vlined,linesnumbered]{algorithm2e}
\usepackage{pgf}
\usepackage{tikz}
\usepackage{pgffor}
\usepackage{stmaryrd}
\usepackage{todonotes}

\usetikzlibrary{decorations.pathmorphing}
\usetikzlibrary{decorations.shapes}
\usetikzlibrary{shapes}
\usetikzlibrary{petri}
\usetikzlibrary{automata}
\usetikzlibrary{backgrounds}

\usetikzlibrary{shadows}
\usetikzlibrary{positioning}

\usetikzlibrary{fit}
\usetikzlibrary{arrows}
\usetikzlibrary{calc}

\usetikzlibrary{matrix}
\tikzstyle{state}=[draw,minimum size=7mm,circle,fill=blue!10]
\tikzstyle{zstate}=[draw,minimum size=5mm,circle,fill=blue!10,font=\footnotesize]
\tikzstyle{ostate}=[draw,minimum size=5mm,rectangle,fill=green!10,font=\footnotesize]
\tikzstyle{pnode}=[draw,minimum size=1mm,circle,fill=red!70!black,text=white,font=\scriptsize,inner sep=1pt]
\tikzstyle{textnode}=[font=\footnotesize]
\tikzstyle{ttextnode}=[font=\tiny]
\tikzstyle{tedge}=[->,line width=0.5pt]
\tikzstyle{pedge}=[->,line width=0.5pt]
\tikzstyle{problemnode}=[fill=green!10,rounded corners,font=\tiny,draw]

\tikzset{background rectangle/.style={fill=yellow!10,rounded corners}}

\newcommand{\blue}[1]{\textcolor{blue}{#1}}
\newcommand{\ignore}[1]{}

\newcommand{\sizeof}[1]{|{#1}|}

\newcommand{\restrictedto}[2]{{#2}|{#1}}
\newcommand{\fun}{f}

\newcommand{\aset}{A}
\newcommand{\bset}{B}
\newcommand{\complementof}[1]{\overline{#1}}
\newcommand{\leqdef}{\sqsubseteq}

\newcommand{\numdefof}[1]{\sizeof{#1}}
\newcommand{\instantiationsof}[1]{\left\llbracket {#1} \right\rrbracket}
\newcommand{\cinstantiationsof}[2]{\left\llbracket {#2} \right\rrbracket_{#1}}
\newcommand{\nat}{{\mathbb N}}
\newcommand{\intgrs}{{\mathbb Z}}

\newcommand{\tuple}[1]{\left\langle{#1}\right\rangle}
\newcommand{\set}[1]{\left\{#1\right\}}
\newcommand{\setcomp}[2]{\left\{{#1}|\;{#2}\right\}}
\newcommand{\assigned}{\leftarrow}
\newcommand{\undef}{\bot}
\newcommand{\domof}[1]{{\it dom}\left({#1}\right)}
\newcommand{\union}{\oplus}
\newcommand{\similar}{\sim}
\newcommand{\game}{{\mathcal G}}
\newcommand{\stateset}{Q}

\newcommand{\istateset}{\stateset_i}
\newcommand{\zstateset}{\stateset_0}
\newcommand{\ostateset}{\stateset_1}
\newcommand{\transitionset}{T}
\newcommand{\gametuple}{\tuple{\stateset,\transitionset,\coloring}}
\newcommand{\alphabet}{\Sigma}
\newcommand{\actionlabel}{\lambda}
\newcommand{\vasstuple}{\tuple{\stateset,\transitionset,\alphabet,\actionlabel}}
\newcommand{\unlabvasstuple}{\tuple{\stateset,\transitionset}}
\newcommand{\fsstate}{s}
\newcommand{\fsstates}{S}
\newcommand{\fstuple}{\tuple{\fsstates,\movesto{a},\alphabet}}
\newcommand{\cntrset}{{\mathcal C}}
\newcommand{\cntrs}{C}
\newcommand{\cntr}{c}
\newcommand{\coloring}{\kappa}
\newcommand{\coloringof}[1]{\coloring\left({#1}\right)}
\newcommand{\col}{{\tt color}}
\newcommand{\conf}{\gamma}

\newcommand{\confs}{\beta}
\newcommand{\confset}{\Gamma}
\newcommand{\intconfset}{\Theta}
\newcommand{\pconfset}[1]{\confset^{#1}}
\newcommand{\zconfset}{\confset_0}
\newcommand{\oconfset}{\confset_1}
\newcommand{\iconfset}{\confset_i}
\newcommand{\pintconfset}[1]{\intconfset^{#1}}

\newcommand{\ipconfset}[1]{\confset^{#1}_i}
\newcommand{\zpintconfset}[1]{\intconfset^{#1}_0}
\newcommand{\opintconfset}[1]{\intconfset^{#1}_1}
\newcommand{\ipintconfset}[1]{\intconfset^{#1}_i}

\newcommand{\cntrval}{\vartheta}

\newcommand{\stateof}[1]{{\tt state}\left({#1}\right)}
\newcommand{\cntrvalof}[1]{{\tt val}\left({#1}\right)}
\newcommand{\state}{q}

\newcommand{\transition}{t}
\newcommand{\op}{{\it op}}
\newcommand{\nop}{{\it nop}}
\newcommand{\inc}[1]{{#1}\mbox{\small + \hspace{-2mm} +}}
\newcommand{\dec}[1]{{#1}\mbox{\small - \hspace{-1mm} -}}
\newcommand{\sourceof}[1]{{\tt source}\left(#1\right)}
\newcommand{\targetof}[1]{{\tt target}\left(#1\right)}
\newcommand{\opof}[1]{{\tt op}\left(#1\right)}
\newcommand{\movesto}[1]{\stackrel{#1}\longrightarrow}
\newcommand{\energy}{{\mathcal E}}
\newcommand{\vass}{{\mathcal V}}
\newcommand{\sem}{{\tt sem}}
\newcommand{\emovesto}[1]{\stackrel{#1}\longrightarrow_\energy}
\newcommand{\vmovesto}[1]{\stackrel{#1}\longrightarrow_\vass}
\newcommand{\semmovesto}[1]{\stackrel{#1}\longrightarrow_\sem}
\newcommand{\path}{\pi}

\newcommand{\run}{\rho}
\newcommand{\runfrom}[3]{{\tt run}\left(#1,#2,#3\right)}

\newcommand{\eventually}{\Diamond}
\newcommand{\always}{\Box}
\newcommand{\strat}{\sigma}

\newcommand{\istrat}{\strat_i}
\newcommand{\notistrat}{\strat_{1-i}}
\newcommand{\zstrat}{\strat_0}
\newcommand{\ostrat}{\strat_1}
\newcommand{\stratset}{\Sigma}
\newcommand{\semistratset}{\stratset_i^\sem}
\newcommand{\semnotistratset}{\stratset_{1-i}^\sem}
\newcommand{\semzstratset}{\stratset_0^\sem}
\newcommand{\semostratset}{\stratset_1^\sem}
\newcommand{\vasszstratset}{\stratset_0^\vass}
\newcommand{\vassostratset}{\stratset_1^\vass}
\newcommand{\energyzstratset}{\stratset_0^\energy}
\newcommand{\energyostratset}{\stratset_1^\energy}

\newcommand{\formula}{\phi}
\newcommand{\winset}{{\mathcal W}}

\newcommand{\ucof}[1]{{#1}\!\uparrow}
\newcommand{\dcof}[1]{{#1}\!\downarrow}
\newcommand{\uclosed}{U}
\newcommand{\ordering}{\preceq}

\newcommand{\sordering}{\prec}

\newcommand{\labeling}{\lambda}
\newcommand{\labelingof}[1]{\labeling\left({#1}\right)}

\newcommand{\toexplore}{{\tt ToExplore}}

\newcommand{\create}[1]{{\tt new}\left({#1}\right)}

\newcommand{\minofthis}[1]{{\it min}\left(#1\right)}
\newcommand{\maxofthis}[1]{{\it max}\left(#1\right)}

\newcommand{\coveredby}{\lhd}

\newcommand{\nonneg}{\overline{\tt neg}}
\newcommand{\parity}{{\tt Parity}}
\newcommand{\pareto}{{\tt Pareto}}
\newcommand{\losingstate}{\state_\ell}
\newcommand{\outgame}{\game^{\it out}}
\newcommand{\outconf}{\conf^{\it out}}
\newcommand{\outstate}{\state^{\it out}}
\newcommand{\outgametuple}{\tuple{\outstateset,\outtransitionset,\outcoloring}}
\newcommand{\outstateset}{\stateset^{\it out}}
\newcommand{\zoutstateset}{\zstateset^{\it out}}
\newcommand{\ooutstateset}{\ostateset^{\it out}}
\newcommand{\outtransitionset}{\transitionset^{\it out}}
\newcommand{\outcoloring}{\coloring^{\it out}}
\newcommand{\outcoloringof}[1]{\outcoloring\left({#1}\right)}

\newcommand{\decpath}[1]{\alpha(#1)}
\newcommand{\incpath}[1]{\overline\alpha(#1)}
\newcommand{\betapath}[1]{\beta(#1)}

\newcommand{\minpump}{v}
\newcommand{\minstart}{u}

\newcommand{\posmucalcul}{L^{\textit{pos}}_\mu}
\newcommand{\guardmucalcul}{L^{\textit{sv}}_\mu}
\newcommand{\var}{X}
\newcommand{\varset}{\mathcal{X}}
\newcommand{\least}{\mu}
\newcommand{\greatest}{\nu}
\newcommand{\env}{\rho}
\newcommand{\interpretationsof}[1]{\llbracket #1 \rrbracket}
\newcommand{\formulabis}{\psi}
\newcommand{\func}{G}
\newcommand{\avass}{\mathcal{S}}

\newcommand{\subformulae}[1]{\mathit{sub}(#1)}
\newcommand{\freevarsof}[1]{\mathit{free}(#1)}
\newcommand{\condition}{{\tt Cond}}

\newcommand{\parg}[1]{\vspace{-2mm}\paragraph{#1}}

\newcommand{\winsetpara}[2]{\winset[{#1}]({#2})}
\newcommand{\paretopara}[2]{\pareto[{#1}]({#2})}

\begin{document}

\maketitle
\begin{abstract}
We consider parity games on infinite graphs where configurations 
are represented by control-states and integer vectors. 
This framework subsumes two classic game problems: parity games on
vector addition systems with states ({\sc vass}) and multidimensional
energy parity games.
We show that the multidimensional energy parity game problem
is inter-reducible with a subclass of single-sided parity games on {\sc vass}
where just one player can modify the integer counters and the
opponent can only change control-states.
Our main result is that the minimal elements of the upward-closed
winning set of these single-sided parity games on {\sc vass}
are computable. This implies that the Pareto frontier of the minimal  
initial credit needed to win multidimensional energy parity games
is also computable, solving an open question from
the literature.
Moreover, our main result implies the decidability of weak 
simulation preorder/equivalence between finite-state systems and 
{\sc vass}, and the decidability of model checking {\sc vass}
with a large fragment of the modal $\mu$-calculus.
\end{abstract}

\section{Introduction}
In this paper, we consider {\it integer games}:
two-player turn-based games
where a color (natural number) is associated
to each state, and where
the transitions allow incrementing and decrementing the values
of a finite set of integer-valued counters by constants.
We refer to the players as Player $0$ and Player $1$.

We consider the classical parity condition, together with two different semantics for integer games:
the {\em energy semantics} and the {\sc vass} {\em semantics}.
The former corresponds to 
{\it multidimensional energy parity games} \cite{CRR:CONCUR2012}, and the latter to
parity games on {\sc vass} (a model essentially equivalent to Petri nets 
\cite{esparza-decidability-94}).
In energy parity games, the winning objective for Player $0$ combines
a qualitative property, the classical {\it parity condition},
with a quantitative property, namely the
{\it energy condition}.
The latter means that the values
of all counters stay above a finite threshold
during the entire run of the game.
In {\sc vass} parity games, the counter values 
are restricted to {\it natural numbers}, and in particular
any transition that may decrease the value of a counter
below zero is disabled (unlike in energy games where such a transition would be
immediately winning for Player 1). So for {\sc vass} games, the objective
consists only of a parity condition, since the energy condition
is trivially satisfied.

We formulate and solve our problems using a generalized
notion of game configurations, namely 
{\it partial configurations},
in which only a subset $\cntrs$ of the counters
may be defined.
A partial configuration $\conf$ denotes
a (possibly infinite) set of concrete configurations that are
called {\it instantiations} of $\conf$.
A configuration $\conf'$ is an instantiation of $\conf$
if $\conf'$ agrees with $\conf$ on the values of the counters
in $\cntrs$ while the values of counters outside $\cntrs$
can be chosen freely in $\conf'$.
We declare a partial configuration to be {\it winning}
(for Player $0$) if  it has an instantiation that is winning.
For each decision problem and each set of counters
$\cntrs$,  we will consider the {\it $\cntrs$-version} of the problem
where we reason about
configurations in which the counters in $\cntrs$ are defined.

\parg{Previous Work.}
Two special cases of the general $\cntrs$-version are
the {\it abstract} version in which no counters are defined, and the
{\it concrete} version in which all counters are defined.
In the energy semantics, the abstract version corresponds to
the {\it unknown initial credit problem} for multidimensional
energy parity games, which is coNP-complete \cite{Chatterjee:FSTTCS2010,CRR:CONCUR2012}.
The concrete version corresponds to the {\it fixed initial credit problem}.
For energy games without the parity condition, the fixed initial 
credit problem was solved in \cite{icalp:BrazdilJK10}
(although it does not explicitly mention energy games but 
instead formulates the problem as a {\it zero-reachability}
objective for Player $1$).
It follows from \cite{icalp:BrazdilJK10} that the
fixed initial credit problem for $d$-dimensional energy games 
can be solved in $d$-EXPTIME (resp. $(d-1)$-EXPTIME for
offsets encoded in unary) and even the upward-closed winning 
sets can be computed. 
An EXPSPACE lower bound is derived by a reduction from
Petri net coverability.
The subcase of one-dimensional energy parity games was
considered in \cite{Chatterjee-Doyen:TCS2012},
where both the unknown and fixed initial credit problems are decidable,
and the winning sets (i.e., the minimal required initial energy) can be computed.
The assumption of having just one dimension is
an important restriction that significantly simplifies the problem.
This case is solved using an algorithm which is
a generalization of the classical algorithms of
McNaughton \cite{McNaughton:APL1993}
and
Zielonka \cite{zielonka1998infinite}.

However, for general multidimensional
energy parity games, computing the winning sets was an open problem,
mentioned, e.g., in \cite{Chatterjee:FSTTCS2010}.

In contrast, under the {\sc vass} semantics,
all these integer game problems are shown to be undecidable
for dimensions $\ge 2$ in \cite{vass:game:CSL}, even for simple
safety/coverability objectives. (The one-dimensional case is a
special case of parity games on one-counter machines, which is
PSPACE-complete).
A special subcase are {\it single-sided} {\sc vass} games,
where just Player $0$ can modify counters while Player $1$
can only change control-states. This restriction makes the winning set
for Player $0$ upward-closed, unlike in general {\sc vass} games.
The paper \cite{raskin-games-05} shows decidability
of coverability objectives for single-sided {\sc vass} games,
using a standard backward fixpoint computation.
\ignore{
It mentions that single-sided games correspond to models where
Player $0$ describes the behavior and the decisions
taken by the system,
while Player $1$ describes the environment in which the
system is embedded.
In many practical cases, the behavior of
the environment can be captured by  a finite-state system.
}

\parg{Our Contribution.}
First we show how instances of the single-sided {\sc vass}
parity game can be reduced to the multidimensional energy parity game, 
and vice-versa. I.e., energy games correspond to the single-sided subcase of 
{\sc vass} games.
Notice that, since parity conditions are closed under complement, it is merely a 
convention that Player $0$ (and not Player $1$) is the one that can change the counters.

Our main result is the decidability of single-sided {\sc vass}
parity games for general partial configurations, and thus in particular 
for the concrete and abstract versions described above.
The winning set for Player $0$ is upward-closed
(wrt.\ the natural multiset ordering on configurations),
and it can be computed 
(i.e., its finitely many minimal elements).
Our algorithm uses the Valk-Jantzen construction \cite{valk-residue-85} 
and a technique similar to Karp-Miller graphs, and finally
reduces the problem to instances of the 
abstract parity problem under the energy semantics, i.e.,
to the unknown initial credit problem in multidimensional energy parity games,
which is decidable by \cite{CRR:CONCUR2012}.

From the above connection between single-sided {\sc vass}
parity games and multidimensional energy parity games,
it follows that the winning sets of 
multidimensional energy parity games are also computable.
I.e., one can compute the {\it Pareto frontier}
of the minimal initial energy credit vectors required to win the 
energy parity game. This solves the problem left open
in \cite{Chatterjee:FSTTCS2010,CRR:CONCUR2012}.

Our results imply further decidability results in the following two areas:
semantic equivalence checking and model-checking.
\ignore{
In \cite{Parosh:Bengt:Karlis:Tsay:general} decidability
results were shown for {\it strong} simulation preorder between
{\sc vass} and finite-state systems, but the decidability of weak 
simulation was left open for one direction.
}
Weak simulation preorder between a finite-state system and a general
{\sc vass} can be reduced to a parity game on a single-sided {\sc vass},
and is therefore decidable. Combined with the previously known decidability of the
reverse direction \cite{Parosh:Bengt:Karlis:Tsay:general},
this implies decidability of weak simulation equivalence.
This contrasts with the undecidability of weak {\em bisimulation} equivalence
between {\sc vass} and finite-state systems \cite{JEM:JCSS1999}.
%
The model-checking problem for {\sc vass} is decidable for 
many linear-time temporal logics \cite{habermehl-complexity-97},
but undecidable even for very restricted branching-time logics \cite{esparza-decidability-94}. 
We show the decidability of model-checking 
for a restricted class of {\sc vass} with a large fragment of the modal
 $\mu$-calculus. 
Namely we consider {\sc vass} where some states do not
perform any updates on the counters, and these states are used to guard the
for-all-successors modal operators in this fragment of the $\mu$-calculus,
allowing us to reduce the model-checking problem to a parity game on single-sided
{\sc vass}.

\section{Integer Games}
\label{igames:section}

\parg{Preliminaries.}
We use $\nat$ and $\intgrs$
to denote the sets of natural numbers (including $0$) and integers respectively.
For a set $\aset$, we define $\sizeof\aset$ to be the cardinality of $\aset$.
For a function $\fun:\aset\mapsto\bset$ from a
set $\aset$ to a set $\bset$, we use
$\fun[a\assigned b]$ to denote the function
$\fun'$ such that $\fun(a)=b$ and $\fun'(a')=\fun(a')$ if $a'\neq a$.
If $f$ is partial, then $\fun(a)=\undef$ means that $\fun$ is undefined for $a$.
In particular $\fun[a\assigned\undef]$ makes the value
of $a$ undefined.
We define $\domof{\fun}:=\setcomp{a}{\fun(a)\neq\undef}$.
%


\parg{Model.}
We assume a finite set $\cntrset$ of {\it counters}.
An {\it integer game} is a tuple
$\game=\gametuple$ where
$\stateset$ is a finite set of {\it states},
$\transitionset$ is a finite set of {\it transitions},
and
$\coloring:\stateset\mapsto\set{0,1,2,\ldots,k}$ is a {\it coloring}
function that assigns to each $\state\in\stateset$
a natural number in the interval $[0..k]$ for
some pre-defined $k$.
The set $\stateset$ is partitioned into two sets
$\zstateset$ (states of Player $0$) and
$\ostateset$ (states of Player $1$).
A transition $\transition\in\transitionset$
is a triple $\tuple{\state_1,\op,\state_2}$
where $\state_1,\state_2\in\stateset$ are states and
$\op$ is an operation of one of the following three forms
(where $\cntr\in\cntrset$ is a counter):
(i) $\inc\cntr$ increments the value of $\cntr$ by one;
(ii) $\dec\cntr$ decrements the value of $\cntr$ by one;
(iii) $\nop$ does not change the value of any counter.
We define
$\sourceof\transition=\state_1$,
$\targetof\transition=\state_2$, and
$\opof\transition=\op$.
We say that $\game$ is {\it single-sided}
in case $\op=\nop$ for all transitions $\transition\in\transitionset$
with $\sourceof\transition\in\ostateset$.
In other words, in a single-sided game,
Player $1$ is not allowed to changes the values of the counters,
but only the state.

\parg{Partial Configurations.}
A {\it partial counter valuation}
$\cntrval:\cntrset\mapsto\intgrs$ is a partial function from the set
of counters to $\intgrs$.
We also write $\cntrval(\cntr)=\undef$ if $\cntr \notin \domof{\cntrval}$.
A {\it partial configuration} $\conf$ is a pair
$\tuple{\state,\cntrval}$ where
$\state\in\stateset$ is a state
and $\cntrval$ is a partial counter valuation. 
We will also consider {\it nonnegative partial configurations},
where the partial counter valuation 
takes values in $\nat$ instead of $\intgrs$.
We define $\stateof\conf:=\state$,
$\cntrvalof\conf:=\cntrval$,
 and
$\coloringof\conf:=\coloringof{\stateof\conf}$.
We generalize assignments from counter valuations to configurations
by defining $\tuple{\state,\cntrval}[\cntr\assigned x] 
= \tuple{\state,\cntrval[\cntr\assigned x]}$.
Similarly, for a configuration
$\conf$ and $\cntr \in \cntrset$ we let 
$\conf(\cntr) := \cntrvalof\conf(\cntr)$,
$\domof\conf:=\domof{\cntrvalof{\conf}}$ and 
$\sizeof\conf:=\sizeof{\domof\conf}$.
For a set of counters $\cntrs\subseteq\cntrset$,
we define
$\pintconfset\cntrs:=\setcomp{\conf}{\domof\conf=\cntrs}$, i.e.,
it is the set of configurations in which the defined counters
are exactly those in $\cntrs$. We use $\pconfset\cntrs$ 
to denote the restriction of $\pintconfset\cntrs$ to nonnegative partial configurations.
We partition $\pintconfset\cntrs$ into two sets
$\zpintconfset\cntrs$ (configurations belonging to Player $0$)
and $\opintconfset\cntrs$ (configurations belonging to Player $1$),
such that $\conf \in \ipintconfset\cntrs$
iff $\domof{\conf}=\cntrs$ and $\stateof\conf\in\istateset$
for $i\in\set{0,1}$.
A configuration is {\it concrete} if
$\domof\conf=\cntrset$, i.e., $\conf\in\pintconfset\cntrset$
(the counter valuation
$\cntrvalof\conf$ is defined for all counters);
and it is {\it abstract} if
$\domof\conf=\emptyset$, i.e., $\conf\in\pintconfset\emptyset$
(the counter valuation
$\cntrvalof\conf$ is not defined for any counter).
 In the sequel, we occasionally write $\intconfset$ instead of $\pintconfset\cntrset$,
and $\intconfset_i$ instead of $\ipintconfset\cntrset$
for $i\in\set{0,1}$. 
The same notations are defined over nonnegative partial configurations with $\confset$, and $\ipconfset\cntrs$ and $\iconfset$ for $i \in \set{0,1}$.
For a nonnegative partial configuration $\conf = \tuple{\state,\cntrval} \in \confset$, and 
set of counters $\cntrs\subseteq\cntrset$
we define the restriction of $\conf$ to $\cntrs$ by 
$\conf' = \restrictedto{\cntrs}{\conf}
= \tuple{\state',\cntrval'}$ where
$\state' = \state$ and 
$\cntrval'(\cntr) = \cntrval(\cntr)$ if $\cntr \in \cntrs$ and
$\cntrval'(\cntr) = \undef$ otherwise.

\parg{Energy Semantics.}
Under the energy semantics, an integer game
induces a transition relation $\emovesto{}$
on the set of partial configurations as follows.
For partial configurations
$\conf_1=\tuple{\state_1,\cntrval_1}$,
$\conf_2=\tuple{\state_2,\cntrval_2}$,
and
a transition $\transition=\tuple{\state_1,\op,\state_2}\in\transitionset$, 
we have
$\conf_1\emovesto{\transition}\conf_2$ 
if one of the following three cases is satisfied:
(i) $\op=\inc\cntr$ and either both
$\cntrval_1(\cntr)=\undef$ and $\cntrval_2(\cntr)=\undef$ or
$\cntrval_1(\cntr)\neq\undef$, $\cntrval_2(\cntr)\neq\undef$ and
$\cntrval_2=\cntrval_1[\cntr\assigned\cntrval_1(\cntr)+1]$;
(ii) $\op=\dec\cntr$, and either both
$\cntrval_1(\cntr)=\undef$ and $\cntrval_2(\cntr)=\undef$ or
$\cntrval_1(\cntr)\neq\undef$, $\cntrval_2(\cntr)\neq\undef$ and
$\cntrval_2=\cntrval_1[\cntr\assigned\cntrval_1(\cntr)-1]$;
(iii) $\op=\nop$ and $\cntrval_2=\cntrval_1$.
Hence we apply the operation
of the transition only if the relevant 
counter value is defined (otherwise, the counter
remains undefined).
Notice that, for a partial configuration $\conf_1$ and a transition
$\transition$, there is at most one $\conf_2$
with $\conf_1\emovesto{\transition}\conf_2$.
If such a $\conf_2$ exists, we define $\transition(\conf_1):=\conf_2$;
otherwise we define $\transition(\conf_1):=\undef$.
We say that $\transition$ is {\it enabled} at $\conf$
if $\transition(\conf)\neq\undef$.
We observe that, in the case of energy semantics,
$\transition$ is not enabled only if
$\stateof\conf\neq\sourceof\transition$.
%
%

\parg{VASS Semantics.}
The difference between the energy and {\sc vass} semantics is
that counters in the case of {\sc vass} range over the natural numbers
(rather than the integers), i.e. the {\sc vass} semantics will be interpreted over nonnegative partial configurations.
Thus,
the transition relation $\vmovesto{}$ induced by an integer game
$\game=\gametuple$ under the {\sc vass} semantics
differs from the one induced by the energy semantics in the sense that
counters are not allowed to assume negative
values. Hence  $\vmovesto{}$  is the restriction of $\emovesto{}$ to nonnegative partial configurations.
Here, a transition $\transition=\tuple{\state_1,\dec\cntr,\state_2}\in\transitionset$ is enabled from $\conf_1=\tuple{\state_1,\cntrval_1}$ only if
 $\cntrval_1(\cntr)>0$ or $\cntrval_1(\cntr)=\undef$.
We assume without restriction that at least
one transition is enabled from each partial configuration
(i.e., there are no deadlocks) in the {\sc vass} semantics
(and hence also in the energy semantics).
%
%
Below, we use $\sem\in\set{\energy,\vass}$ to distinguish the
energy and {\sc vass} semantics.

\parg{Runs.}
A {\it run} $\run$ in semantics $\sem$ is an infinite sequence
$\conf_0\semmovesto{\transition_1}
\conf_1\semmovesto{\transition_2}
\cdots$ of transitions between concrete configurations.
A {\it path} $\path$ in $\sem$ is a finite sequence
$\conf_0\semmovesto{\transition_1}
\conf_1\semmovesto{\transition_2}
\cdots\conf_n$ of transitions between concrete configurations.
We say that $\run$ (resp.\ $\path$) is
a $\conf$-run (resp.\ $\conf$-path) if
$\conf_0=\conf$.
We define $\run(i):=\conf_i$ and $\path(i):=\conf_i$.
We assume familiarity with the logic LTL.
For an LTL formula $\formula$ we write $\run\models_\game\formula$
to denote that the run $\run$ in $\game$ satisfies $\formula$.
For instance,
given a set $\confs$ of concrete configurations,
we write $\run\models_\game\eventually\confs$ to denote that
there is an $i$ with
$\conf_i\in\confs$ (i.e., a member of $\confs$
eventually occurs  along $\run$);
and
write $\run\models_\game\always\eventually\confs$ to denote that
there are infinitely many $i$ with
$\conf_i\in\confs$ (i.e., members of $\confs$ occur infinitely
often along $\run$).

\parg{Strategies.}
A {\it strategy} of Player $i\in\set{0,1}$ in $\sem$
(or simply an $i$-strategy in $\sem$) $\istrat$ is a mapping
that assigns to each path
$\path=\conf_0\semmovesto{\transition_1}
\conf_1\semmovesto{\transition_2}
\cdots\conf_n$ with $\stateof{\conf_n}\in\istateset$,
a transition $\transition=\istrat(\path)$ with
$\transition(\conf_n)\neq\undef$ in $\sem$.
We use $\semistratset$  to denote the sets of
$i$-strategies in $\sem$.
Given a concrete configuration $\conf$,
$\zstrat\in\semzstratset$, and
$\ostrat\in\semostratset$, we
define $\runfrom\conf{\zstrat}{\ostrat}$ to be the unique run
$\conf_0\semmovesto{\transition_1}
\conf_1\semmovesto{\transition_2}
\cdots$
such that
(i) $\conf_0=\conf$,
(ii) $\transition_{i+1}=\zstrat(\conf_0\semmovesto{\transition_1}
\conf_1\semmovesto{\transition_2}
\cdots\conf_i)$ if $\stateof{\conf_i}\in\zstateset$, and
(iii) $\transition_{i+1}=\ostrat(\conf_0\semmovesto{\transition_1}
\conf_1\semmovesto{\transition_2}
\cdots\conf_i)$ if $\stateof{\conf_i}\in\ostateset$.
For $\istrat\in\semistratset$,
we write $[i,\istrat,\sem]:\conf\models_\game\formula$ to denote that
$\runfrom\conf{\istrat}{\notistrat}\models_\game\formula$
for all $\notistrat\in\semnotistratset$.
In other words, Player $i$ has a {\it winning strategy},
namely $\istrat$, which ensures that $\formula$ will
be satisfied regardless of the strategy chosen by Player $1-i$.
We write $[i,\sem]:\conf\models_\game\formula$ to denote that
$[i,\istrat,\sem]:\conf\models_\game\formula$ for some
$\istrat\in\semistratset$.

\parg{Instantiations.}
Two nonnegative partial configurations $\conf_1,\conf_2$ are said to be {\it disjoint}
if
(i)  $\stateof{\conf_1}=\stateof{\conf_2}$, and
(ii) $\domof{\conf_1}\cap\domof{\conf_2}=\emptyset$
(notice that we require the states to be equal).
For a set of counters $\cntrs\subseteq\cntrset$, and
disjoint partial configurations $\conf_1,\conf_2$,
we say that $\conf_2$ is a {\it $\cntrs$-complement}
of $\conf_1$ if $\domof{\conf_1}\cup\domof{\conf_2}=\cntrs$,
i.e.,  $\domof{\conf_1}$ and $\domof{\conf_2}$ form
a partitioning of the set $\cntrs$.
If $\conf_1$ and $\conf_2$ are disjoint then we define
$\conf_1\union\conf_2$ to be the nonnegative partial configuration
$\conf:=\tuple{\state,\cntrval}$
such that
$\state:=\stateof{\conf_1}=\stateof{\conf_2}$,
$\cntrval(\cntr):=\cntrvalof{\conf_1}(\cntr)$
if $\cntrvalof{\conf_1}(\cntr)\neq\undef$,
$\cntrval(\cntr):=\cntrvalof{\conf_2}(\cntr)$
if $\cntrvalof{\conf_2}(\cntr)\neq\undef$, and
$\cntrval(\cntr):=\undef$ if both
$\cntrvalof{\conf_1}(\cntr)=\undef$ and
$\cntrvalof{\conf_2}(\cntr)=\undef$.
In such a case, we say that $\conf$ is a {\it $\cntrs$-instantiation}
of $\conf_1$.
For a nonnegative partial configuration $\conf$
we write
$\cinstantiationsof\cntrs\conf$
to denote the set
of $\cntrs$-instantiations of $\conf$.
We will consider the special case where
$\cntrs=\cntrset$.
In particular,
we say that $\conf_2$ is  a {\it complement} of
$\conf_1$ if
$\conf_2$ is a $\cntrset$-complement
of $\conf_1$, i.e.,
 $\stateof{\conf_2}=\stateof{\conf_1}$ and
$\domof{\conf_1}=\cntrset-\domof{\conf_2}$.
We use $\complementof\conf$ to denote the set of complements
of $\conf$.
If $\conf_2\in\complementof{\conf_1}$,
we say that $\conf=\conf_1\union\conf_2$ is an {\it instantiation}
of $\conf_1$.
Notice that $\conf$ in such a case is concrete.
For a nonnegative partial configuration $\conf$
we write
$\instantiationsof\conf$
to denote the set
of instantiations of $\conf$.
We observe that $\instantiationsof\conf=\cinstantiationsof\cntrset\conf$
and that $\instantiationsof\conf=\set{\conf}$
for any concrete nonnegative configuration $\conf$.

\parg{Ordering.}
For nonnegative partial configurations $\conf_1,\conf_2$,
we write $\conf_1\similar\conf_2$ if
$\stateof{\conf_1}=\stateof{\conf_2}$ and
$\domof{\conf_1}=\domof{\conf_2}$.
We write
$\conf_1\leqdef\conf_2$ if
$\stateof{\conf_1}=\stateof{\conf_2}$ and
$\domof{\conf_1}\subseteq\domof{\conf_2}$.
For nonnegative partial
configurations
$\conf_1\similar\conf_2$,
we write
$\conf_1\ordering\conf_2$ to denote that
$\stateof{\conf_1}=\stateof{\conf_2}$ and
$\cntrvalof{\conf_1}(\cntr)\leq\cntrvalof{\conf_2}(\cntr)$
for all $\cntr\in\domof{\conf_1}=\domof{\conf_2}$.
For a nonnegative partial configuration
$\conf$,
we define $\ucof\conf:=\setcomp{\conf'}{\conf\ordering\conf'}$
to be the upward closure of $\conf$,
and define  $\dcof\conf:=\setcomp{\conf'}{\conf'\ordering\conf}$
to be the downward closure of $\conf$.
Notice that $\ucof\conf=\dcof\conf=\set{\conf}$
for any abstract configuration $\conf$.
For a set $\confs\subseteq\pconfset\cntrs$ of nonnegative partial configurations, let
$\ucof\confs:=\cup_{\conf\in\confs}\ucof\conf$.
We say that $\confs$ is upward-closed if $\ucof\confs=\confs$.
For an upward-closed set $\confs\subseteq\pconfset\cntrs$, we use
$\minofthis\confs$ to denote the (by Dickson's Lemma unique and finite)
set of minimal elements of $\confs$.

\parg{Winning Sets of Partial Configurations.}
For  a nonnegative partial configuration $\conf$,
we write
$[i,\sem]:\conf\models_\game\formula$ to denote that
$\exists\conf'\in\instantiationsof\conf.[i,\sem]:\conf'\models_\game\formula$,
i.e., Player $i$ is winning from some instantiation
$\conf'$ of $\conf$.
For a set $\cntrs\subseteq\cntrset$ of counters,
we define
$\winsetpara{\game,\sem,i,\cntrs}{\formula}:=
\setcomp{\conf\in\pconfset\cntrs}{[\sem,i]:\conf\models_\game\formula}$.
If $\winsetpara{\game,\sem,i,\cntrs}{\formula}$ is upward-closed,
we define the {\it Pareto frontier} as
$\paretopara{\game,\sem,i,\cntrs}{\formula}:=
\minofthis{\winsetpara{\game,\sem,i,\cntrs}{\formula}}$.

\parg{Properties.}
We show some useful properties of the ordering
on nonnegative partial configurations. 
Note that for nonnegative partial configurations, we will not make distinctions between the energy semantics and the {\sc vass} semantics; 
this is due to the fact that in nonnegative partial configurations and in their instantiations we only consider positive values for the counters. 
For the energy semantics, as we shall see, this will not be a problem since we will consider winning runs where the counter never goes below $0$. 
We now show {\it monotonicity}
and (under some conditions) ``reverse monotonicity''  of
the transition relation wrt. $\ordering$.
%
%
%
We write $\conf_1\semmovesto{}\conf_2$ if there exists $\transition$ such that $\conf_1\semmovesto{\transition}\conf_2$.  
\begin{lemma}
\label{monotonicity:lemma}
Let  $\conf_1$, $\conf_2$, and $\conf_3$ be nonnegative partial configurations.
If
(i)
$\conf_1\vmovesto{}\conf_2$, and
(ii)
$\conf_1\ordering\conf_3$,
then there is a
$\conf_4$ such that
$\conf_3\vmovesto{}\conf_4$ and
$\conf_2\ordering\conf_4$.
Furthermore, if
(i)
$\conf_1\vmovesto{}\conf_2$, and
(ii)
$\conf_3\ordering\conf_1$, and
(iii)
$\game$ is single-sided and
(iv) $\conf_1 \in\oconfset$,
then there is a
$\conf_4$ such that
$\conf_3 \vmovesto{}\conf_4$ and
$\conf_4\ordering\conf_2$.
\end{lemma}
We consider a version of the {\it Valk-Jantzen}
lemma \cite{valk-residue-85}, expressed in our terminology.

\begin{lemma}\cite{valk-residue-85}
\label{vj:lemma}
Let $\cntrs\subseteq\cntrset$ and let
$\uclosed\subseteq\pconfset\cntrs$
be upward-closed.
Then, $\minofthis\uclosed$ is computable if and only if,
for any nonnegative partial configuration $\conf$ with $\domof{\conf} \subseteq \cntrs$, we can decide whether
$\cinstantiationsof\cntrs{\conf}\cap\uclosed\neq\emptyset$.
\end{lemma}
%
%
%

\section{Game Problems}
\label{problems:section}

\begin{wrapfigure}{R}{.6\textwidth}
\begin{tikzpicture}[show background rectangle]
\node[name=dummy]{};
\node at (dummy)[problemnode,name=absenergy,align=center]
{Abstract Energy \\ \blue{decidable \cite{CRR:CONCUR2012}}};

\node at ($(absenergy)+(27mm,0mm)$) [problemnode,name=cvass,align=center]
{$\cntrs$-version \\ Single-Sided {\sc vass}\\ \blue{decidable, Corollary~\ref{corollary:vass:decidable}}};

\node at ($(cvass)+(27mm,0mm)$) [problemnode,name=concretevass,align=center]
{Concrete \\ Single-Sided {\sc vass}\\ \blue{decidable}};

\node at ($(cvass)+(0mm,-13mm)$) [problemnode,name=cenergy,align=center]
{$\cntrs$-version  Energy\\ \blue{decidable, Corollary~\ref{corollary:energy:decidable}}};

\node at ($(cenergy)+(27mm,0mm)$) [problemnode,name=concreteenergy,align=center]
{Concrete Energy\\ \blue{decidable}};

\node at ($(cvass)+(0mm,13mm)$) [problemnode,name=paretovass,align=center]
{Pareto\\Single-Sided {\sc vass}\\ \blue{computable, Theorem~\ref{single:sided:vass:parity:theorem}}};

\node at ($(paretovass)+(27mm,0mm)$) [problemnode,name=paretoenergy,align=center]
{Pareto Energy\\ \blue{computable, Theorem~\ref{parity:energy:vass:theorem}}};

\draw[pedge] (cvass) to node[ttextnode,above=-1pt]{Algorithm~\ref{graph:algorithm}} (absenergy) ;
\draw[pedge] (cenergy.36) to node[ttextnode,right=-1pt]{Lemma~\ref{lem:energy-vass}} (cvass.315) ;
\draw[pedge] (cvass.225) to node[ttextnode,left=-1pt]{Lemma~\ref{conf:parity:vass:energy:lemma}} (cenergy.143) ;
\draw[pedge] (concretevass) to node[ttextnode,above=-1pt]{Trivial} (cvass) ;
\draw[pedge] (concreteenergy) to node[ttextnode,above=-1pt]{Trivial} (cenergy) ;
\draw[pedge] (paretovass) to node[ttextnode,above,sloped]{Section~\ref{one:sided:parity:section}} (absenergy) ;
\draw[pedge] (paretovass) to node[ttextnode,right=-1pt]{Section~\ref{one:sided:parity:section}} (cvass) ;
\draw[pedge] (paretoenergy.west) to node[ttextnode,above=-1pt]{Lemma~\ref{lem:energy-vass}} (paretovass.east)  ;

\end{tikzpicture}
\caption{Problems considered in the paper and their relations.
For each property, we state the lemma that show its decidability/computability.
The arrows show the reductions of problem instances
that we show in the paper.}
\label{problems:fig}
\end{wrapfigure}
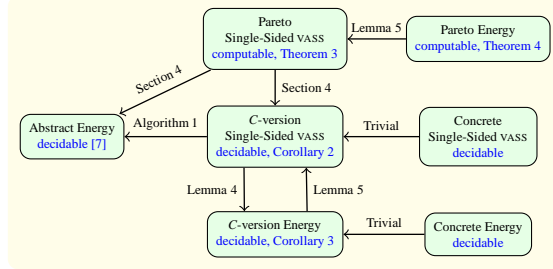
Here we consider the parity winning condition for the integer
games defined in the previous section.
First we establish a correspondence between the
{\sc vass} semantics when the underlying integer
game is single-sided, and the energy semantics in the general case.
We will show how instances of the single-sided {\sc vass}
parity game can be reduced to the energy parity game, and vice-versa.
Figure~\ref{problems:fig} depicts 
a summary of
our results.
For either semantics, an instance of the problem consists of an integer game
$\game$ and a partial configuration $\conf$.
For a given set of counters
$\cntrs\subseteq\cntrset$, we will consider
the {\it $\cntrs$-version} of the problem where we
assume that $\domof\conf=\cntrs$.
In particular, we will consider two special cases:
(i) the {\it abstract} version in which we  assume
that $\conf$ is abstract (i.e., $\domof\conf=\emptyset$), and
(ii) the {\it concrete} version in which
we  assume that $\conf$ is concrete
(i.e., $\domof\conf=\cntrset$).
The abstract version of a problem corresponds to
the {\it unknown initial credit problem} \cite{Chatterjee:FSTTCS2010,CRR:CONCUR2012}, 
while the concrete one
corresponds to deciding if a given initial credit is sufficient
or, more generally, computing the Pareto frontier
(left open in \cite{Chatterjee:FSTTCS2010,CRR:CONCUR2012}).

\parg{Winning Conditions.}
Assume an integer game $\game=\gametuple$
where $\coloring:\stateset\mapsto\set{0,1,2,\ldots,k}$.
For a partial configuration $\conf$ and $i:0\leq i\leq k$,
the relation $\conf\models_\game(\col=i)$ holds
if $\coloringof{\stateof\conf}=i$.
The formula simply checks the color of the state of $\conf$.
The formula $\conf\models_\game\nonneg$ holds
if $\cntrvalof\conf(\cntr)\geq 0$
for all $\cntr\in\domof\conf$.
The formula states that the values of all counters are nonnegative
in $\conf$.
For $i:0\leq i\leq k$, 
the predicate ${\it even}(i)$ holds if $i$ is even.
Define the path formula
$\parity:=
\bigvee_{(0\leq i\leq k)\wedge {\it even}(i)}
\left(
\left(\always\eventually(\col=i)\right)
\wedge
\left(
\bigwedge_{i<j\leq k}\eventually\always \neg(\col=j)
\right)
\right)
$.
The formula states that the highest color that appears infinitely often
along the path is even.

\parg{\it Energy Parity.}
Given an integer game $\game$ and a partial
configuration $\conf$, we ask whether
$[0,\energy]:\conf\models_\game\parity\wedge(\always\nonneg)$, i.e.,
whether Player $0$ can force a run
in the energy semantics
where the parity condition is satisfied and
at the same time the counters remain nonnegative.
The abstract version of this problem is equivalent to the
unknown initial credit problem in classical
energy parity games \cite{Chatterjee:FSTTCS2010,CRR:CONCUR2012}, since it amounts to asking for
the {\em existence} of a threshold for the initial counter values from which Player $0$ can win.
%
The nonnegativity objective $(\always\nonneg)$ justifies our restriction to
nonnegative partial configurations in our definition of the instantiations 
and hence of the winning sets.

\begin{theorem}\cite{CRR:CONCUR2012}
\label{abstract:energy:theorem}
The abstract energy parity problem is decidable.
\end{theorem}

The winning set $\winsetpara{\game,\energy,0,\cntrs}{\parity\wedge\always\nonneg}$
is upward-closed for $\cntrs\subseteq\cntrset$. 
Intuitively, if Player $0$ can win the game with a certain value for the
counters, then any higher value for these counters also allows him to win the
game with the same strategy. This is because both the possible moves of Player
$1$ and the colors of configurations depend only on the control-states.

\begin{lemma}
\label{energy:parity:uc:lemma}
For any $\cntrs\subseteq\cntrset$, the set $\winsetpara{\game,\energy,0,\cntrs}{\parity\wedge\always\nonneg}$ is upward-closed.
\end{lemma}

Since this winning set is upward-closed, it follows from Dickson's Lemma that
it has finitely many minimal elements. These minimal elements describe the
Pareto frontier of the minimal initial credit needed to win the game. 
In the sequel we will show how to compute this set 
$\paretopara{\game,\energy,0,\cntrs}{\parity\wedge\always\nonneg)}
:= \minofthis{\winsetpara{\game,\energy,0,\cntrs}{\parity\wedge\always\nonneg}}$;
cf. Theorem~\ref{parity:energy:vass:theorem}.

\parg{VASS Parity.}
Given an integer game $\game$ and a nonnegative partial
configuration $\conf$, we ask whether
$[0,\vass]:\conf\models_\game\parity$, i.e.,
whether Player $0$ can force a run
in the {\sc vass} semantics
where the parity condition is satisfied.
(The condition $\always\nonneg$ is always trivially satisfied in {\sc vass}.)
In general, this problem is undecidable as shown in \cite{vass:game:CSL},
even for simple coverability 
 objectives instead of parity objectives.
\begin{theorem}\cite{vass:game:CSL}
\label{vass:undecidability:theorem}
The {\it VASS Parity Problem} is undecidable.
\end{theorem}

We will show that decidability of the {\sc vass} parity problem is regained under the assumption
that $\game$ is single-sided. In \cite{raskin-games-05} it was already shown
that, for a single-sided {\sc vass} game with
reachability objectives, it is possible to compute the set of winning
configurations.
However, the proof for parity objectives is much more involved.

\parg{Correspondence of Single-Sided {\sc vass} Games and Energy Games.}
We show that single-sided {\sc vass} parity games can be reduced 
to energy parity games, and vice-versa.
The following lemma shows the direction from {\sc vass} to energy.

\begin{lemma}\label{conf:parity:vass:energy:lemma}
Let $\game$ be a single-sided integer game and let $\conf$ be a nonnegative partial configuration.
Then $[0,\vass]:\conf\models_\game\parity$
iff
$[0,\energy]:\conf\models_\game\parity\wedge\always\nonneg$.
\end{lemma}

Hence for a single-sided $\game$ and any set 
$\cntrs\subseteq\cntrset$, we have
$\winsetpara{\game,\vass,0,\cntrs}{\parity}=\winsetpara{\game,\energy,0,\cntrs}{\parity\wedge\always\nonneg}$. 
Consequently,
using Lemma~\ref{energy:parity:uc:lemma} and
Theorem~\ref{abstract:energy:theorem},
we obtain the following corollary.

\begin{corollary}\label{cor:vass-energy} 
Let $\game$ be single-sided and $\cntrs\subseteq\cntrset$. 
\begin{enumerate}
\item
$\winsetpara{\game,\vass,0,\cntrs}{\parity}$ is upward-closed.
\item
The $\cntrs$-version single-sided {\sc vass} parity problem is reducible
to the $\cntrs$-version energy parity problem.
\item
The {\em abstract} single-sided {\sc vass} parity problem (i.e., where
$\cntrs=\emptyset$)
is decidable.
\end{enumerate}
\end{corollary}

The following lemma shows the reverse reduction from energy parity games to
single-sided {\sc vass} parity games.

\begin{lemma}\label{lem:energy-vass}
Given an integer game $\game=\gametuple$,
one can construct a single-sided integer game
$\game' = \tuple{\stateset',\transitionset',\coloring'}$ 
with $\stateset \subseteq \stateset'$
such that 
$[0,\energy]:\conf\models_\game\parity\wedge\always\nonneg$
iff
$[0,\vass]:\conf\models_{\game'}\parity$
for every nonnegative partial configuration $\conf$ of $\game$.
\end{lemma}
{\em Proof sketch.}
%
Since $\game'$ needs to be single-sided, Player $1$ cannot change the
counters. Thus the construction forces Player $0$ to simulate the moves of
Player $1$. Whenever a counter drops below zero in $\game$ (and thus Player
$0$ loses), Player $0$ cannot perform this simulation in $\game'$ and is forced to go to a losing state instead.
\qed

\parg{Computability Results.} 
The following theorem (shown in
Section~\ref{one:sided:parity:section})
states our main computability result.
For single-sided {\sc vass} parity games, the 
minimal elements of the winning set 
$\winsetpara{\game,\vass,0,\cntrs}{\parity}$ (i.e., the Pareto frontier) 
are computable.

\begin{theorem}
\label{single:sided:vass:parity:theorem}
If $\game$ is single-sided then
$\paretopara{\game,\vass,0,\cntrs}{\parity}$ is computable.
\end{theorem}

In particular, this implies decidability.
\begin{corollary}
\label{corollary:vass:decidable}
For any set of counters $\cntrs\subseteq\cntrset$, the $\cntrs$-version single-sided {\sc vass} parity problem is decidable.
\end{corollary}

From Theorem~\ref{single:sided:vass:parity:theorem} and
Lemma~\ref{lem:energy-vass} we obtain the computability of the Pareto frontier
of the minimal initial credit needed to win general energy parity games.

\begin{theorem}
\label{parity:energy:vass:theorem}
$\paretopara{\game,\energy,0,\cntrs}{\parity \wedge\always\nonneg}$ is computable for any game $\game$.
\end{theorem}

\begin{corollary}
\label{corollary:energy:decidable}
The $\cntrs$-version energy parity problem is decidable.
\end{corollary}

\section{Solving Single-Sided VASS Parity Games (Proof of Theorem~\ref{single:sided:vass:parity:theorem})}
\label{one:sided:parity:section}
%
Consider a single-sided integer game $\game=\gametuple$ and
a set $\cntrs\subseteq\cntrset$ of counters.
We will show how to compute the set
$\paretopara{\game,\vass,0,\cntrs}{\parity}$.
We reduce the problem of computing the Pareto frontier
in the single-sided {\sc vass} parity game
to solving the abstract energy parity game problem, which is decidable
by Theorem~\ref{abstract:energy:theorem}.

We use induction on $k=\sizeof\cntrs$.
As we shall see, the base case is straightforward.
We perform the induction step in two phases.
First
we show that, under the induction hypothesis,
we can reduce the problem of computing the
Pareto frontier to the problem of solving the
$\cntrs$-version single-sided {\sc vass} parity problem
(i.e., we need only to consider individual nonnegative partial configurations
in $\pconfset\cntrs$).
In the second phase, we introduce an algorithm that
translates the latter problem to the abstract
energy parity problem.

\parg{Base Case.}
Assume that $\cntrs=\emptyset$.
In this case we are considering the abstract single-sided
{\sc vass} parity problem.
Recall that $\ucof{\conf}=\set{\conf}$
for any $\conf$ with $\domof\conf=\emptyset$.
Since $\cntrs=\emptyset$, it follows that
$\paretopara{\game,\vass,0,\cntrs}{\parity}=
\setcomp{\conf}{(\domof\conf=\emptyset)\wedge
\left([0,\vass]:\conf\models_\game\parity\right)}$.
In other words, computing the Pareto frontier in this case reduces
to solving the  abstract single-sided {\sc vass} parity problem,
which is decidable by Corollary~\ref{cor:vass-energy}.

\parg{From Pareto Sets to {\sc vass} Parity.}
Assuming the induction hypothesis,
we reduce the problem of computing the  set
$\paretopara{\game,\vass,0,\cntrs}{\parity}$ to
the $\cntrs$-version single-sided {\sc vass} parity problem, i.e.,
the problem of checking whether $[0,\vass]:\conf\models_\game\parity$
for some $\conf\in\pconfset\cntrs$ when the underlying integer
game is single-sided.
To do that, we will instantiate the Valk-Jantzen lemma as follows.
We instantiate $\uclosed \subseteq \confset^\cntrs$ in Lemma~\ref{vj:lemma}
to be $\winsetpara{\game,\vass,0,\cntrs}{\parity}$
(this set is upward-closed by Corollary~\ref{cor:vass-energy}
since $\game$ is single-sided). Take any nonnegative partial configuration
$\conf$ with $\domof{\conf} \subseteq \cntrs$. We consider two cases.
First, if $\domof{\conf}=\cntrs$, then we are dealing with the $\cntrs$-version
single-sided {\sc vass} parity game
which will show how to solve in the sequel.
Second, consider the case where $\domof{\conf}=\cntrs' \subset \cntrs$. By the induction hypothesis, we can compute the (finite) set
$\paretopara{\game,\vass,0,\cntrs'}{\parity}=\minofthis{\winsetpara{\game,\vass,0,\cntrs'}{\parity}}$. Then to solve this case, we use the following lemma.
\begin{lemma}
\label{lem:subconf}
For all nonnegative partial configurations $\conf$ such that $\domof{\conf}=\cntrs' \subset \cntrs$, we have $\cinstantiationsof\cntrs{\conf}\cap
\winsetpara{\game,\vass,0,\cntrs}{\parity}\neq\emptyset$ iff $\conf \in \winsetpara{\game,\vass,0,\cntrs'}{\parity}$.
\end{lemma}
Hence checking $\cinstantiationsof\cntrs{\conf}\cap
\winsetpara{\game,\vass,0,\cntrs}{\parity}\neq\emptyset$ amounts to simply
comparing $\conf$ with the elements of the finite set
$\paretopara{\game,\vass,0,\cntrs'}{\parity}$,
because $\winsetpara{\game,\vass,0,\cntrs'}{\parity}$ is upward-closed by
Corollary~\ref{cor:vass-energy}.

\parg{From {\sc vass} Parity to Abstract Energy Parity.}
We introduce an algorithm that
uses the induction hypothesis to translate an instance
of the $\cntrs$-version single-sided {\sc vass} parity problem
to an equivalent instance of the abstract energy parity problem.

The following definition and lemma
formalize some consequences of the induction hypothesis.
First we define a relation that allows us to directly classify some
nonnegative partial configurations as winning for Player $1$ (resp. Player $0$).

\begin{definition}\label{def:coveredby}
Consider a nonnegative partial configuration
$\conf$ and a set
of nonnegative partial configurations $\confs$.
We write $\confs\coveredby\conf$ if:
(i) for each $\hat{\conf}\in\confs$,
$\domof{\hat{\conf}} \subseteq \cntrs$
and $\numdefof{\conf}=\numdefof{\hat{\conf}}+1$, and
(ii) for each $\cntr\in\domof\conf$ there
is a $\hat{\conf}\in\confs$
such that
$\hat{\conf}\ordering\conf[\cntr\assigned\undef]$.
\end{definition}

\begin{lemma}\label{lem:coveredby}
Let
$\confs = \bigcup_{\cntrs' \subseteq \cntrs, |\cntrs'|=|\cntrs|-1}
\paretopara{\game,\vass,0,\cntrs'}{\parity}$
be the Pareto frontier of minimal Player $0$ winning nonnegative partial configurations with
one counter in $\cntrs$ undefined.
Let $\{c_i, \dots, c_j\} = \cntrset - \cntrs$ be the counters outside $\cntrs$.
%
\begin{enumerate}
\item
For every $\hat{\conf} \in \confs$ with
$\{\cntr\} = \cntrs - \domof{\hat{\conf}}$
there exists a minimal finite number $\minpump(\hat{\conf})$
s.t. $\instantiationsof{\hat{\conf}[\cntr\assigned \minpump(\hat{\conf})]} \cap
{\winsetpara{\game,\vass,0,\cntrset}{\parity}} \neq \emptyset$.
\item
For every $\hat{\conf} \in \confs$ there is a number $\minstart(\hat{\conf})$
s.t. $\hat{\conf}[\cntr\assigned \minpump(\hat{\conf})][\cntr_i \assigned
  \minstart(\hat{\conf}), \dots, \cntr_j \assigned \minstart(\hat{\conf})]
\in {\winsetpara{\game,\vass,0,\cntrset}{\parity}}$,
i.e., assigning value $\minstart(\hat{\conf})$ to counters outside $\cntrs$ is sufficient to make the
nonnegative configuration winning for Player $0$.
\item
If $\conf \in\pconfset\cntrs$ is a Player $0$
winning nonnegative partial configuration, i.e.,
$\instantiationsof{\conf} \cap {\winsetpara{\game,\vass,0,\cntrset}{\parity}} \neq
\emptyset$, then $\confs\coveredby\conf$.
\end{enumerate}
\end{lemma}

The third part of this lemma implies that if $\neg(\confs\coveredby\conf)$
then we can directly conclude that $\conf$ is not winning for Player $0$ (and
thus winning for Player $1$) in the parity game.

Now we are ready to present the algorithm
(Algorithm~\ref{graph:algorithm}).

\parg{Input and output of the algorithm.}
The algorithm
inputs a single-sided integer game $\game=\gametuple$,
and a nonnegative partial configuration $\conf$
where $\domof\conf=\cntrs$.
To check whether $[0,\vass]:\conf\models_\game\parity$,
it constructs an instance of the abstract energy parity problem.
This instance is defined by a new integer game
$\outgame=\tuple{\stateset_{\it out},\transitionset_{\it out},\coloring_{\it out}}$
with counters in $\cntrset - \cntrs$,
and a nonnegative partial configuration $\outconf$.
Since we are considering the abstract version of the problem,
the configuration $\outconf$ is
of the form $\outconf=\tuple{\outstate,\cntrval_{\it out}}$
where $\domof{\cntrval_{\it out}}=\emptyset$.
The latter property means that $\outconf$ is uniquely
determined by the state $\outstate$
(all counter values are undefined).
Lemma~\ref{algorithm:pcorrectness:lemma} relates
$\game$ with the newly constructed $\outgame$.

\begin{algorithm}
{\scriptsize
\KwIn{
$\game=\gametuple$: Single-Sided Integer Game;
\hspace{3mm}
$\conf\in\pconfset\cntrs$ with $\sizeof\cntrs=k>0$.
}
\KwOut{$\outgame=\outgametuple$:
integer game;\\
\hspace{11mm}
$\outstate\in\outstateset$; $\outconf=\tuple{\outstate,\cntrval_{\it out}}$
where $\domof{\cntrval_{\it out}}=\emptyset$;
\hspace{2mm}
$\labeling:\stateset_{\it out}\cup\transitionset_{\it out}\mapsto\;\pconfset\cntrs\cup\transitionset$
}

$\confs\gets
\bigcup_{(\cntrs'\subseteq\cntrs)\wedge\sizeof{\cntrs'}=\sizeof\cntrs-1}
\paretopara{\game,\vass,0,\cntrset'}{\parity}$
\label{pareto:init:line}\;

$\;\outtransitionset\gets\emptyset$;\hspace{2mm}
$\create\outstate$;\hspace{2mm}
$\coloringof\outstate\gets\coloringof\conf$;\hspace{2mm}
$\labelingof\outstate\gets\conf$;\hspace{2mm}
$\stateset_{\it out}\gets\set{\outstate}$\label{graph:init:line}\;
\lIf{$\labelingof\outstate\in\zconfset$}{$\zoutstateset\gets\set{\outstate}$;\hspace{2mm}$\ooutstateset\gets\emptyset$}\hspace{2mm}\lElse{$\ooutstateset\gets\set{\outstate}$;\hspace{2mm}$\zoutstateset\gets\emptyset$} \label{init:zstateset:line}\;
$\toexplore\gets\set{\outstate}$
\label{toexplore:init:line}\;

\While{$\toexplore\neq\emptyset$\label{while:line}}{
  Pick and remove a $\state\in\toexplore$\label{pick:line}\;
  \If{$\neg(\confs\coveredby\labelingof\state)$\label{lose:line}}{
    $\outcoloringof\state\gets 1$;\hspace{2mm}
    $\outtransitionset\gets\outtransitionset\cup\set{\tuple{\state,\nop,\state}}$
  }
  \lElse{\If{$\exists\state'.\left(\state',\state\right)\in\left(\outtransitionset\right)^*\wedge \left(\labeling(\state')\sordering\labelingof\state\right)$ \label{win:line}}
    { $\outcoloringof\state\gets 0$;\hspace{2mm}
      $\outtransitionset\gets\outtransitionset\cup\set{\tuple{\state,\nop,\state}}$
    }
       \lElse{\For{each $\transition\in\transitionset$ with $\transition(\labelingof\state)\neq\undef$\label{for:line}}{
           \If{$\exists\state'.\left(\state',\state\right)\in\left(\outtransitionset\right)^*.\labelingof{\state'}=\transition(\labelingof\state)$\label{copy:line}}{
             $\outtransitionset\gets\outtransitionset\cup\set{\tuple{\state,\opof\transition,\state'}}$;\hspace{2mm}
             $\labelingof{\tuple{\state,\opof\transition,\state'}}\gets\transition$
             }
             \Else{
               $\create{\state'}$;\hspace{2mm} \label{new:state:line}
               $\coloringof{\state'}\gets\coloringof{\transition(\labelingof\state)}$;\hspace{2mm}
               $\labeling(\state')\gets\transition(\labelingof\state)$\;
              \lIf{$\labelingof{\state'}\in\zconfset$}{$\zoutstateset\gets\zoutstateset\cup\set{\state'}$}\hspace{2mm}\lElse{$\ooutstateset\gets\ooutstateset\cup\set{\state'}$} \label{zstateset:line}\;
               $\outtransitionset\gets\outtransitionset\cup\set{\tuple{\state,\opof\transition,\state'}}$\label{add:to:transitionset:line};\hspace{2mm}
               $\labeling(\tuple{\state,\opof\transition,\state'})\gets\transition$\;
               $\toexplore\gets\toexplore\cup\set{\state'}$; \label{add:to:explore:line}
             }
         }
       }
  }
}

}
\caption{Building an instance of the abstract energy parity problem.}
\label{graph:algorithm}
\end{algorithm}

\parg{Operation of the algorithm.}
The algorithm performs a forward analysis similar to the
classical Karp-Miller algorithm for Petri nets. We
start with a given nonnegative partial configuration, explore its successors,
create loops when previously visited configurations are repeated and
define a special operation for the case when configurations strictly increase.
The algorithm builds  the graph of the game $\outgame$
successively
(i.e., the set of states $\outstateset$,
the set of transitions $\outtransitionset$, and the coloring
of states $\coloring$).
Additionally, for bookkeeping purposes inside the algorithm
and for reasoning about the correctness of the algorithm,
we define a labeling function $\labeling$
on the set of states and transitions in $\outgame$
such that each state in $\outgame$ is labeled by a nonnegative partial configuration
in $\pconfset\cntrs$, and each transition in $\outgame$
is labeled by a transition in $\game$.

The algorithm first computes the Pareto frontier
$\paretopara{\game,\vass,0,\cntrs'}{\parity}$ for
all counter sets $\cntrs'\subseteq\cntrset$
with $\sizeof{\cntrs'}=\sizeof{\cntrs}-1$.
This is possible by the induction hypothesis.
It stores the union of all these sets in $\confs$ (line~\ref{pareto:init:line}).
At line~\ref{graph:init:line}, the algorithms initializes the
 set of transitions $\outtransitionset$ to be empty,
creates the first state $\outstate$, defines its coloring
to be the same as that of the state of the input nonnegative partial configuration $\conf$,
labels it by  $\conf$, and then adds it to the set of
states $\outstateset$.
At line~\ref{init:zstateset:line} it adds  $\outstate$ to the set
of states of Player $0$ or Player $1$ (depending on where $\conf$ belongs), and
at line~\ref{toexplore:init:line} it adds $\outstate$ to the set $\toexplore$.
The latter contains the set of states that have been created but
not yet analyzed by the algorithm.

After the initialization phase, the algorithm starts iterating the
{\it while}-loop starting at line~\ref{while:line}.
During each iteration, it picks and removes a new state $\state$
from the set $\toexplore$ (line~\ref{pick:line}).
First, it checks two special conditions under which the game is made
immediately losing (resp. winning) for Player $0$.

{\bf Condition 1:}
If $\neg(\confs\coveredby\labelingof\state)$ (line~\ref{lose:line}),
then we know by Lemma~\ref{lem:coveredby} (item 3) that the nonnegative partial configuration
$\labelingof\state$ is not winning
for Player $0$ in $\game$.

Therefore, we make the state $\state$ losing
for Player $0$ in $\outgame$.
To do that,
we change the color of $\state$ to $1$ (any odd color will do),
and add a self-loop to $\state$.
Any continuation of a run from $\state$ is then losing
for Player $0$ in $\outgame$.

{\bf Condition 2:}
If Condition 1 did not hold then the algorithm checks (at line~\ref{win:line})
whether there is a {\em predecessor} $\state'$ of $\state$ in $\outgame$
with a label $\labelingof{\state'}$ that is {\it strictly}
smaller than the label $\labelingof{\state}$
of $\state$, i.e., $\labelingof{\state'}\sordering\labelingof\state$.
(Note that we are not comparing $\state$ to arbitrary other states in $\outgame$,
but only to predecessors.)
If that is the case, then the state $\state$ is made winning for Player $0$
in $\outgame$.
To do that,
we change the color of $\state$ to $0$ (any even color will do),
and add a self-loop to $\state$.
The intuition for making $\state$ winning for Player $0$ is as follows.
Since $\labelingof{\state'}\sordering\labelingof\state$,
the path from $\labelingof{\state'}$ to
$\labelingof{\state}$  increases the value of at least
one of the defined counters (those in $\cntrs$), and will not decrease the
other counters in $\cntrs$ (though it might have a negative effect on the
undefined counters in $\cntrset -\cntrs$).
Thus, if a run in $\game$
iterates this path sufficiently many times, the value
of at least one counter in $\cntrs$ will be pumped and
becomes sufficiently high to allow Player $0$ to win the parity game on
$\game$, provided that the counters in $\cntrset -\cntrs$ are initially
instantiated with sufficiently high values. This follows from
the property $\confs\coveredby\labelingof\state$ and
Lemma~\ref{lem:coveredby} (items 1 and 2).

\ignore{
Later we will show that the introduction of Condition 2 does not unfairly prevent Player
$1$ from winning.
}

If none of the tests for Condition1/Condition2 at lines~\ref{lose:line}~and~\ref{win:line}
succeeds, the algorithm continues expanding the graph
of $\outgame$ from $\state$.
It generates all successors of $\state$ by applying
each transition $\transition\in\transitionset$ in $\game$
to the label $\labelingof\state$ of $\state$ (line~\ref{for:line}).
If the result $\transition(\labelingof\state)$
is defined then there are two possible cases.
The first case occurs if we have previously encountered (and added to
$\outstateset$) a state $\state'$ whose label
equals $\transition(\labelingof\state)$ (line~\ref{copy:line}).
Then we add a transition from $\state$ back to
$\state'$ in $\outgame$, where the operation of the new
transition is the same operation as that of $\transition$,
and define the label of the new transition to be $\transition$.
Otherwise (line~\ref{new:state:line}), we create a new state $\state'$,
label it with the nonnegative configuration $\transition(\labelingof{\state})$
and assign it the same color as  $\transition(\labelingof{\state})$.
At line~\ref{zstateset:line} $\outstate$ is added to the set
of states of Player $0$ or Player $1$ (depending on where $\conf$ belongs).
We add a new transition between $\state$ and $\state'$
 with the same operation as $\transition$.
The new transition is labeled with $\transition$.
Finally, we add the new state $\state'$ to the set of states to be explored.

\begin{lemma}
\label{algorithm:termination:lemma}
Algorithm~\ref{graph:algorithm} will always terminate.
\end{lemma}

Lemma~\ref{algorithm:termination:lemma} implies that
the integer game $\outgame$ is finite (and hence well-defined).
The following lemma shows the relation between the
input and output games $\game,\outgame$.

\begin{lemma}
\label{algorithm:pcorrectness:lemma}
$[0,\vass]:\conf\models_\game\parity$
iff
$[0,\energy]:\outconf\models_{\outgame}\parity\wedge\always\nonneg$ .
\end{lemma}
{\em Proof sketch.}
The left to right implication is easy. Given a Player $0$ winning strategy in
$\game$, one can construct a winning strategy in $\outgame$ that uses the same
transitions, modulo the labeling function $\labelingof{}$.
The condition $\always\nonneg$ in $\outgame$ is satisfied since
the configurations in $\game$ are always nonnegative and the parity condition
is satisfied since the colors seen in corresponding plays in $\outgame$ and $\game$ are the same.

For the right to left implication we consider a Player $0$ winning strategy
$\zstrat$ in $\outgame$ and construct a winning strategy $\zstrat'$ in
$\game$. The idea is that a play $\path$ in $\game$ induces a play $\path'$ in
$\outgame$ by using the same sequence of transitions, but removing all
so-called {\em pumping sequences}, which are subsequences that
end in {\bf Condition 2}. Then $\zstrat'$ acts on history $\path$
like $\zstrat$ on history $\path'$.
For a play according to $\zstrat'$ there are two cases.
Either it will eventually reach a configuration that is
sufficiently large (relative to $\confs$)
such that a winning strategy is known by induction hypothesis.
Otherwise it contains only finitely many pumping sequences and an infinite
suffix of it coincides with an infinite suffix of a play according to
$\zstrat$ in $\outgame$. Thus it sees the same colors and satisfies
$\parity$. \qed

Since $\outconf$ is abstract and
the abstract energy parity problem is decidable
(Theorem~\ref{abstract:energy:theorem})
we obtain Theorem~\ref{single:sided:vass:parity:theorem}.

The termination proof in Lemma~\ref{algorithm:termination:lemma}
relies on Dickson's Lemma, and thus there is no elementary upper
bound on the complexity of Algorithm~\ref{graph:algorithm}
or on the size of the constructed energy game $\outgame$.
The algorithm in \cite{icalp:BrazdilJK10} for the fixed initial credit problem
in pure energy games without the parity condition runs in $d$-exponential time
(resp. $(d-1)$-exponential time for offsets encoded in unary) for dimension $d$,
and is thus not elementary either.
As noted in \cite{icalp:BrazdilJK10}, the best known lower bound is EXPSPACE
hardness, easily obtained via a reduction from the control-state reachability (i.e., coverability)
problem for Petri nets.

\section{Applications to Other Problems}

\subsection{Weak simulation preorder between VASS and finite-state systems}

Weak simulation preorder \cite{Gla2001} is a semantic preorder on the states of labeled
transition graphs, which can be characterized by weak simulation games.
A configuration of the game is given by a pair of states $(\state_1,\state_0)$.
In every round of the game, Player $1$ chooses a labeled step
$\state_1 \movesto{a} \state_1'$ for some label $a$. Then Player $0$ must
respond by a move which is either of the form
$\state_0 \movesto{\tau^*a\tau^*} \state_0'$ if $a \neq \tau$, or
of the form $\state_0 \movesto{\tau^*} \state_0'$ if $a =\tau$ (the special label
$\tau$ is used to model internal transitions).
The game continues from configuration $(\state_1', \state_0')$. A player wins
if the other player cannot move and Player $0$ wins every infinite play.
One says that $\state_0$ {\em weakly simulates} $\state_1$ iff Player $0$ has
a winning strategy in the weak simulation game from $(\state_1,\state_0)$.
States in different transition systems can be compared by putting them
side-by-side and considering them as a single transition system.

We use $\vasstuple$ to denote a labeled {\sc vass} where the states and
transitions are defined as in Section~\ref{igames:section}, $\alphabet$ is a
finite set of labels and $\actionlabel: \transitionset \mapsto \alphabet$
assigns labels to transitions.

It was shown in \cite{Parosh:Bengt:Karlis:Tsay:general} that
it is decidable whether a finite-state labeled transition system
weakly simulates a labeled {\sc vass}.
However, the decidability of the reverse direction was open.
(The problem is that the weak $\movesto{\tau^*a\tau^*}$ moves
in the {\sc vass} make the weak simulation game infinitely branching.)
We now show that it is also decidable whether a labeled {\sc vass}
weakly simulates a finite-state labeled transition system.
In particular this implies that weak simulation equivalence between a
labeled {\sc vass} and a finite-state labeled transition system
is decidable.
This is in contrast to the undecidability of weak {\em bisimulation} equivalence
between {\sc vass} and finite-state systems \cite{JEM:JCSS1999}.

\begin{theorem}\label{thm:weaksim}
It is decidable whether a labeled {\sc vass} weakly simulates a finite-state labeled transition system.
\end{theorem}
{\em Proof sketch.}
Given a labeled {\sc vass} and a finite-state labeled transition system,
one constructs a single-sided {\sc vass} parity game s.t.
the {\sc vass} weakly simulates the finite system iff Player $0$ wins
the parity game. The idea is to take a controlled product of the finite system
and the {\sc vass} s.t. every round of the weak simulation game is encoded by a
single move of Player $1$ followed by an arbitrarily long sequence of moves 
by Player $0$. The move of Player $1$ does not change the counters, since it 
encodes a move in the finite system, and thus the game is single-sided.
Moreover, one enforces that every sequence of consecutive moves by Player $0$ is finite
(though it can be arbitrarily long),
by assigning an odd color to Player $0$ states and a higher even color to Player $1$
states.

\subsection{$\mu$-Calculus model checking VASS}\label{sec:mucalc}

While model checking {\sc vass} with linear-time temporal logics (like LTL
and linear-time $\mu$-calculus) is decidable
\cite{esparza-decidability-94,habermehl-complexity-97},
model checking {\sc vass} with most branching-time logics (like EF, EG, CTL and
the modal $\mu$-calculus) is undecidable \cite{esparza-decidability-94}.
However, we show that Theorem~\ref{single:sided:vass:parity:theorem} yields the
decidability of model checking single-sided {\sc vass} with a guarded fragment
of the modal $\mu$-calculus.
We consider a {\sc vass} $\unlabvasstuple$ where the states, transitions and
semantics are defined as in Section~\ref{igames:section}, and
reuse the notion of partial configurations and the transition relation
defined for the {\sc vass} semantics on integer games.
We specify properties on such {\sc vass} in the
positive $\mu$-calculus $\posmucalcul$ whose atomic propositions $\state$
refer to control-states $\state \in \stateset$ of the input {\sc vass}.

The syntax of the positive $\mu$-calculus $\posmucalcul$ is given
by the following grammar:
$\formula ::=  \state ~\mid~ \var ~\mid~ \formula \wedge \formula ~\mid~
\formula \vee \formula ~\mid~ \eventually \formula ~\mid ~ \always \formula
~\mid~ \least \var.\formula ~\mid~ \greatest \var.\formula$
where $\state \in \stateset$ and $\var$ belongs to a countable set of
variables $\varset$.
The semantics of $\posmucalcul$ is defined as usual (see appendix).
To each closed formula $\formula$ in $\posmucalcul$ (i.e., without free
variables)
it assigns a subset of concrete
configurations $\interpretationsof{\formula}$.

The model-checking problem of {\sc vass} with $\posmucalcul$ can then be
defined as follows.
Given a {\sc vass} $\avass=\unlabvasstuple$, a closed formula $\formula$ of
$\posmucalcul$
and an initial configuration $\conf_0$ of $\avass$, do we have
$\conf_0 \in \interpretationsof{\formula}$? If the answer is
yes, we will write $\avass,\conf_0 \models \formula$.
The more general {\em global model-checking problem} is to compute the set
$\interpretationsof{\formula}$ of configurations that satisfy the
formula. The general unrestricted version of this problem is undecidable.

\begin{theorem}\cite{esparza-decidability-94}
The model-checking problem of {\sc vass} with $\posmucalcul$ is undecidable.
\end{theorem}
One way to solve the $\mu$-calculus model-checking problem for a given Kripke
structure is to encode the problem into a parity game \cite{gradel-automata-02-chapter-9}.
%
The idea is to construct a parity game whose states are pairs, where the first
component is a state of the structure and the second component is a
subformula of the given $\mu$-calculus formula.
States of the form $\tuple{\state,\always \formula}$ or
$\tuple{\state,\formula \wedge \formulabis}$
belong to Player $1$ and the remainder belong to Player $0$.
The colors are assigned to reflect the nesting of least and greatest
fixpoints.
We can adapt this construction to our context by building an integer game from
a formula in $\posmucalcul$ and a {\sc vass} $\avass$, as
stated by the next lemma.

\begin{lemma}
\label{lem:mucalcul:to:parity}
Let $\avass$ be a {\sc vass}, $\conf_0$ a concrete configuration of $\avass$ and $\formula$ a closed formula in $\posmucalcul$. One can construct an integer game $\game(\avass,\formula)$ and an initial concrete configuration $\conf'_0$ such that $[0,\vass]:\conf'_0 \models_{\game(\avass,\formula)}\parity$ if and only if $\avass,\conf_0 \models \formula$.
\end{lemma}

Now we show that, under certain restrictions on the considered {\sc vass}
and on the formula from $\posmucalcul$, the constructed integer game
$\game(\avass,\formula)$ is
single-sided, and hence we obtain the decidability of the model-checking
problem from Theorem~\ref{single:sided:vass:parity:theorem}.
First, we reuse the notion of single-sided games from
Section~\ref{igames:section} in the context of {\sc vass},
by saying that a {\sc vass} $\avass=\unlabvasstuple$ is single-sided iff
there is a partition of the set of states $\stateset$ into two sets $\zstateset$ and
$\ostateset$ such that $\op=\nop$ for all transitions $\transition\in\transitionset$
with $\sourceof\transition\in\ostateset$.
The guarded fragment $\guardmucalcul$ of $\posmucalcul$
for single-sided {\sc vass} is then defined by guarding the $\always$ operator
with a predicate that enforces control-states in $\ostateset$.
Formally, the syntax of $\guardmucalcul$ is given by the following grammar:
$\formula ::=  \state ~\mid~ \var ~\mid~ \formula \wedge \formula ~\mid~
\formula \vee \formula ~\mid~ \eventually \formula ~\mid ~ \ostateset \wedge
\always \formula ~\mid~ \least \var.\formula ~\mid~ \greatest \var.\formula$,
where $\ostateset$ stands for the formula $\bigvee_{\state \in \ostateset} \state$.
By analyzing the construction of Lemma~\ref{lem:mucalcul:to:parity} in this
restricted case, we obtain the following lemma.

\begin{lemma}
\label{lem:mucalcul:single:to:parity}
If $\avass$ is a single-sided {\sc vass} and $\formula \in \guardmucalcul$
then the game $\game(\avass,\formula)$ is equivalent to a single-sided game.
\end{lemma}

By combining the results of the last two lemmas with
Corollary~\ref{cor:vass-energy},
Theorem~\ref{single:sided:vass:parity:theorem} and
Corollary~\ref{corollary:vass:decidable},
we get the following result on model checking single-sided {\sc vass}.

\begin{theorem}~
\label{thm:mucalcul:decidable}
\begin{enumerate}
\item Model checking $\guardmucalcul$ over single-sided {\sc vass} is decidable.
\item If $\avass$ is a single-sided {\sc vass} and $\formula$ is a formula
of $\guardmucalcul$ then $\interpretationsof{\formula}$ is
upward-closed and its set of minimal elements is computable.
\end{enumerate}
\end{theorem}

\section{Conclusion and Outlook}
We have established a connection between multidimensional energy games and
single-sided {\sc vass} games. Thus our algorithm to compute winning sets in 
{\sc vass} parity games can also be used to compute the minimal initial credit
needed to win multidimensional energy parity games, i.e., the Pareto frontier.

It is possible to extend our results to integer parity games with a mixed semantics,
where a subset of the counters follow the energy semantics and the rest
follow the {\sc vass} semantics. 
If such a mixed parity game is single-sided w.r.t. the {\sc vass} counters 
(but not necessarily w.r.t. the energy counters) then it can be reduced to
a single-sided {\sc vass} parity game by our construction in
Section~\ref{problems:section}. The winning set of the derived 
single-sided {\sc vass} parity game
can then be computed with the algorithm in Section~\ref{one:sided:parity:section}.



\newpage 
\section*{Appendix}

\subsection*{Proof of Lemma~\ref{monotonicity:lemma}}

Let  $\conf_1=\tuple{\state_1,\cntrval_1}$, $\conf_2=\tuple{\state_2,\cntrval_2}$, and $\conf_3=\tuple{\state_3,\cntrval_3}$ be nonnegative partial configurations and let $\transition=\tuple{\state'_1,\op,\state'_2}$ in $\transitionset$. 

Assume that $\conf_1 \vmovesto{\transition} \conf_2$ and that
$\conf_1\ordering\conf_3$. From this we know that
$\state_1=\state'_1=\state_3$, that $\state_2=\state'_2$ and also that
$\domof{\conf_1}=\domof{\conf_2}=\domof{\conf_3}$.  There are several cases
for a transition $\transition$ that can be taken from $\conf_1$. 
If $\op$ is an increment or a $\nop$ operation then only the control-state
matters for taking the transition.
If $\op$ is a decrement transition then the initial value of the decremented counter 
has to be either $\undef$ or $\ge 1$. Since this is the case for $\cntrval_1$
and since $\conf_1\ordering\conf_3$, we deduce that this also holds for
$\cntrval_3$. 
Then we obtain the nonnegative partial configuration 
$\conf_4=\tuple{\state_4,\cntrval_4}$ from $\conf_3$ by following the rule 
of the transition relation $\vmovesto{\transition}$. Moreover, 
we can deduce that $\conf_2\ordering\conf_4$, because any operation on the
undefined counters 
leaves the counters undefined, and for the other counters one can easily prove that for all $\cntr' \in \domof{\conf_1}$, $\cntrval_4(\cntr')=\cntrval_2(\cntr')+(\cntrval_3(\cntr')-\cntrval_1(\cntr'))$.

Now suppose that 
$\conf_1\vmovesto{\transition}\conf_2$, that $\conf_3\ordering\conf_1$, that
$\game$ is single-sided and that $\conf_1 \in\oconfset$. It follows that
$\state_1=\state'_1=\state_3$, that $\state_2=\state'_2$ and also that
$\domof{\conf_1}=\domof{\conf_2}=\domof{\conf_3}$. Furthermore, since
$\conf_1 \in\oconfset$, we deduce that $\state_1 \in \ostateset$ and,
since $\game$ is single-sided, we have that $\op=\nop$. 
Hence, by definition of the transition relation $\vmovesto{\transition}$, 
we obtain $\cntrval_1=\cntrval_2$ and so by choosing
$\conf_4=\tuple{\state_2,\cntrval_3}$ we obtain that
$\conf_3\movesto{\transition}\conf_4$. 
Since $\conf_3\ordering\conf_1$, we  have 
$\cntrval_3(\cntr) \leq \cntrval_1(\cntr)$ for all $\cntr \in \domof{\conf_1}$ 
and hence $\conf_4\ordering\conf_2$.

\subsection*{Proof of Lemma~\ref{vj:lemma}}

Usually the Valk and Jantzen Lemma, which allows the computation of the minimal
elements of an upward-closed set of vectors of naturals, is stated a bit
differently by using vectors of naturals with $\omega$ at some indexes to
represent any integer values (see for instance in
\cite{abadulla-minimal-12}). In our context, the $\omega$ are replaced by
undefined values for the counters in the considered nonnegative partial configurations, 
but the idea is the same. 
The usual way to express the Valk and Jantzen Lemma is as follows: 
For $\cntrs\subseteq\cntrset$ and an upward-closed set
$\uclosed\subseteq\pconfset\cntrs$,  
$\minofthis\uclosed$ is computable if and only if
for any nonnegative partial configuration $\conf$ with $\domof{\conf} \subseteq \cntrs$, one can decide whether
$\dcof{\cinstantiationsof\cntrs{\conf}}\cap\uclosed\neq\emptyset$. 
Now we show that this way of formalizing the Valk and Jantzen Lemma 
is equivalent to the statement of Lemma~\ref{vj:lemma}. 

First, if we assume that $\minofthis\uclosed$ is computable, then 
it is obvious that for any nonnegative partial configuration $\conf$ with $\domof{\conf}
\subseteq \cntrs$, 
we can decide whether
$\cinstantiationsof\cntrs{\conf}\cap\uclosed\neq\emptyset$. 
In fact, it suffices to check whether there exists a $\conf_1 \in
\minofthis\uclosed$ 
such that for all $\cntr \in \domof{\conf}$, 
we have $\conf(\cntr) \ge \conf_1(\cntr)$
(since $\uclosed$ is upward-closed). 
Since $\minofthis\uclosed$ is finite, it is possible check this condition for
all nonnegative partial configurations $\conf_1$ in $\minofthis\uclosed$.

Now assume that for any nonnegative partial configuration $\conf$ with $\domof{\conf}
\subseteq \cntrs$, we can decide whether
$\cinstantiationsof\cntrs{\conf}\cap\uclosed\neq\emptyset$. Consider a
configuration $\conf_1$ with $\domof{\conf_1} \subseteq \cntrs$.  First note
that $\dcof{\conf_1}$ is a finite set and also that
$\dcof{\cinstantiationsof\cntrs{\conf_1}}=\bigcup_{\conf_2 \in \dcof{\conf_1}}
\cinstantiationsof\cntrs{\conf_2}$. But since $\dcof{\conf_1}$ is finite, and
since we can decide whether
$\cinstantiationsof\cntrs{\conf_2}\cap\uclosed\neq\emptyset$ for each $\conf_2
\in \dcof{\conf_1}$, we can decide whether
$\dcof{\cinstantiationsof\cntrs{\conf_1}}\cap\uclosed\neq\emptyset$. 
By the Valk and Jantzen Lemma, $\minofthis\uclosed$ is computable.

\subsection*{Proof of Lemma~\ref{energy:parity:uc:lemma}}

We will show that the set $\winsetpara{\game,\energy,0,\cntrs}{\parity\wedge \always\nonneg}$ is upward-closed. Let $\conf_1,\conf_2 \in\pconfset\cntrs$ (with $\conf_1=\tuple{\state_1,\cntrval_1}$ and $\conf_2=\tuple{\state_1,\cntrval_2}$) such that $\conf_1 \in \winsetpara{\game,\energy,0,\cntrs}{\parity\wedge\always\nonneg}$ and $\conf_1 \ordering \conf_2$. In order to prove that $\winsetpara{\game,\energy,0,\cntrs}{\parity\wedge\always\nonneg}$ is upward-closed, we need to show that $\conf_2 \in \winsetpara{\game,\energy,0,\cntrs}{\parity\wedge\always\nonneg}$.

Since $\conf_1 \in \winsetpara{\game,\energy,0,\cntrs}{\parity\wedge\always\nonneg}$, there exists $\conf'_1 \in \instantiationsof{\conf_1}$ such that $[0,\energy]:\conf'_1\models_\game\parity\wedge\always\nonneg$, i.e., there exists $\conf'_1=\tuple{\state_1,\cntrval'_1} \in \instantiationsof{\conf_1}$ and $\zstrat\in\energyzstratset$ such that $\runfrom{\conf'_1}{\zstrat}{\ostrat}\models_\game \parity\wedge\always\nonneg$ for all  $\ostrat\in\energyostratset$. Let us first define the following concrete configuration $\conf'_2=\tuple{\state_1,\cntrval'_2}$ with:

$$
\cntrval'_2(\cntr)=\left\{ \begin{array}{ll}
\cntrval_2(\cntr) & \mbox{ if } \cntr \in \domof{\conf_2} \\
\cntrval'_1(\cntr)  & \mbox{ if } \cntr \notin \domof{\conf_2}\\
\end{array}
\right.
$$
By definition we have $\conf'_2 \in \instantiationsof{\conf_2}$ and since $\conf_1 \ordering \conf_2$, we also have $\conf'_1 \ordering \conf'_2$ (i.e. $\cntrval'_1(\cntr) \leq \cntrval'_2(\cntr)$ for all $\cntr \in \cntrset$). We want to show that $[0,\energy]:\conf'_2 \models_\game \parity\wedge\always\nonneg$, i.e., that player $0$ has a winning strategy from the concrete configuration $\conf'_2$.

We now show how to build a winning strategy $\zstrat' \in \energyzstratset$ for player $0$ from the configuration $\conf'_2$. First to any $\conf'_2$-path $\path=\conf''_0 \emovesto{\transition_1}
\conf''_1\emovesto{\transition_2}
\cdots\conf''_n$ we associate the $\conf'_1$-path $\decpath{\path}=\conf'''_0 \emovesto{\transition_1}
\conf'''_1\emovesto{\transition_2}
\cdots\conf'''_n$ where for all $j \in \set{0,\ldots,n}$, if $\conf''_j=\tuple{\state''_j,\cntrval''_j}$ then $\conf'''_j=\tuple{\state''_j,\cntrval'''_j}$ with $\cntrval'''_j(\cntr)=\cntrval''_j(\cntr)-(\cntrval'_2(\cntr)-\cntrval'_1(\cntr))$ for all $\cntr \in \cntrset$ (i.e. to obtain $\decpath{\path}$ from $\path$, we decrement each counter valuation by the difference between $\cntrval'_2(\cntr)-\cntrval'_1(\cntr)$). Note that $\decpath{\path}$ is a valid path since we are considering the energy semantics where the counters can take negative values. Now we define the strategy $\zstrat' \in \energyzstratset$ for player $0$ as $\zstrat'(\path)=\zstrat(\decpath{\path})$ for each $\conf'_2$-path $\path$. Here again the strategy is well defined since in energy games the enabledness of a transition depends only on the control-state and not on the counter valuation. 

We will now prove that for all strategies $\ostrat' \in \energyostratset$, we
have $\runfrom{\conf'_2}{\zstrat'}{\ostrat'}\models_\game
\parity\wedge\always\nonneg$. Let $\ostrat' \in \energyostratset$.
Using $\ostrat'$, we will construct another strategy $\ostrat \in
\energyostratset$ and prove that if
$\runfrom{\conf'_1}{\zstrat}{\ostrat}\models_\game
\parity\wedge\always\nonneg$ then
$\runfrom{\conf'_2}{\zstrat'}{\ostrat'}\models_\game
\parity\wedge\always\nonneg$. 
Before we give the definition of $\ostrat$, we introduce another notation. To any $\conf'_1$-path $\path=\conf'''_0 \emovesto{\transition_1}
\conf'''_1\emovesto{\transition_2}
\cdots\conf'''_n$ we associate the $\conf'_2$-path $\incpath{\path}=\conf''_0 \emovesto{\transition_1}
\conf''_1\emovesto{\transition_2}
\cdots\conf''_n$ where for all $j \in \set{0,\ldots,n}$, if $\conf'''_j=\tuple{\state''_j,\cntrval'''_j}$ then $\conf''_j=\tuple{\state''_j,\cntrval''_j}$ with $\cntrval''_j(\cntr)=\cntrval'''_j(\cntr)+(\cntrval'_2(\cntr)-\cntrval'_1(\cntr))$ for all $\cntr \in \cntrset$ (i.e., to obtain $\incpath{\path}$ from $\path$, we increment each counter valuation by the difference between $\cntrval'_2(\cntr)-\cntrval'_1(\cntr)$). Note that $\incpath{\path}$ is a valid path. Now we define the strategy $\ostrat \in \energyostratset$ for Player $1$ as $\ostrat(\path)=\ostrat'(\incpath{\path})$ for each $\conf'_1$-path $\path$.

We extend in the obvious way the function $\decpath{}$ [resp. $\incpath{}$] to
$\conf'_2$-run [resp. to $\conf'_1$-run]. Then one can easily check that we
have
$\decpath{\runfrom{\conf'_2}{\zstrat'}{\ostrat'}}=\runfrom{\conf'_1}{\zstrat}{\ostrat}$
and that
$\runfrom{\conf'_2}{\zstrat'}{\ostrat'}=\incpath{\runfrom{\conf'_1}{\zstrat}{\ostrat}}$
by construction of the strategy $\zstrat'$ and $\ostrat$. First, remember that
$\zstrat$ is a winning strategy for Player $0$ from the configuration
$\conf'_1$. Thus we have $\runfrom{\conf'_1}{\zstrat}{\ostrat}\models_\game
\parity\wedge\always\nonneg$. Since in
$\incpath{\runfrom{\conf'_1}{\zstrat}{\ostrat}}$ the sequence of
control-states are the same and all the counter valuations 
along the path are greater or equal to the ones seen in
$\runfrom{\conf'_1}{\zstrat}{\ostrat}$ (remember that we add to each configuration, to each counter $\cntr$ the quantity $\cntrval'_2(\cntr)-\cntrval'_1(\cntr) \geq 0$), this allows us to deduce that $\incpath{\runfrom{\conf'_1}{\zstrat}{\ostrat}} \models_\game \parity\wedge\always\nonneg$. Hence we have $\runfrom{\conf'_2}{\zstrat'}{\ostrat'} \models_\game \parity\wedge\always\nonneg$.

Finally we have proved that there exist $\conf'_2 \in
\instantiationsof{\conf_2}$ and $\zstrat'\in\energyzstratset$ such that
$\runfrom{\conf'_2}{\zstrat'}{\ostrat'}\models_\game
\parity\wedge\always\nonneg$ for all  $\ostrat'\in\energyostratset$. So Player
$0$ has a winning strategy from an instantiation of the configuration
$\conf_2$. Hence $\conf_2 \in \winsetpara{\game,\energy,0,\cntrs}{\parity\wedge\always\nonneg}$.

\subsection*{Proof of Lemma~\ref{conf:parity:vass:energy:lemma}}

Let $\game=\gametuple$ be a single-sided integer game and $\conf\in\pconfset\cntrs$ a nonnegative partial configuration. 

First we will assume that
$[0,\energy]:\conf\models_\game\parity\wedge\always\nonneg$. This means that
there exists $\conf'=\tuple{\state,\cntrval} \in \instantiationsof{\conf}$ and
$\zstrat \in \energyzstratset$ such that
$\runfrom{\conf'}{\zstrat}{\ostrat}\models_\game \parity\wedge\always\nonneg$
for all $\ostrat \in \energyostratset$. 
The idea we will use here
is that since the strategy $\zstrat$ keeps the value of the counters
positive, then the same strategy can be followed under the {\sc vass}
semantics, and furthermore this strategy will be a winning strategy for the
{\sc vass} parity game. Let us formalize this idea. We build the strategy $\zstrat' \in  \vasszstratset$ as follows: for any path $\path=\conf_0 \vmovesto{\transition_1}
\conf_1\vmovesto{\transition_2} \cdots\conf_n$, we have $\zstrat'(\path)=\zstrat(\path)$ if $\zstrat(\path)(\conf_n) \neq  \undef$ (under the {\sc vass} semantics) and otherwise $\zstrat'(\path)$ equals any enabled transition. Note that this definition is valid since any path in the {\sc vass} semantics is also a path in the energy semantics. We consider now a strategy $\ostrat' \in  \vassostratset$. This strategy can be easily extended to a strategy $\ostrat \in \energyostratset$ for the energy game by playing any transition when the input path is not a path valid under the {\sc vass} semantics. First note that since $\zstrat$ is a winning strategy in the energy parity game we have $\runfrom{\conf'}{\zstrat}{\ostrat}\models_\game \parity\wedge\always\nonneg$. From the way we build the strategies, we deduce that $\runfrom{\conf'}{\zstrat'}{\ostrat'}=\runfrom{\conf'}{\zstrat}{\ostrat}$. Since the colors seen along a run depend only of the control-state, we deduce that $\runfrom{\conf'}{\zstrat'}{\ostrat'}\models_\game \parity$. Hence we have proven that  $[0,\vass]:\conf\models_\game\parity$.

We now assume that $[0,\vass]:\conf\models_\game\parity$. This means that there exists $\conf'=\tuple{\state,\cntrval} \in \instantiationsof{\conf}$ and $\zstrat \in \vasszstratset$ such that  $\runfrom{\conf'}{\zstrat}{\ostrat}\models_\game \parity$ for all $\ostrat \in \vassostratset$. We build a strategy $\zstrat' \in \energyzstratset$ as follows: for any path $\path$ in the {\sc vass} semantics $\zstrat'(\path)=\zstrat(\path)$; otherwise, if $\path$ is not a valid path under the {\sc vass} semantics, $\zstrat'(\path)$ is equal to any transition enabled in the last configuration of the path. Take now a strategy $\ostrat' \in \energyzstratset$ for Player 1 in the energy parity game. From $\ostrat'$, we define a strategy $\ostrat \in \vasszstratset$ as follows: for any path $\path$ in the {\sc vass} semantics, let $\ostrat(\path)=\ostrat'(\path)$. Note that since the game is single-sided this strategy is well defined; in fact, in a single-sided game, in the states of Player 1, all the outgoing transitions are enabled in the energy and in the {\sc vass} semantics (because in single-sided games, Player 1 does not change the counter values). But then we have $\runfrom{\conf'}{\zstrat}{\ostrat}=\runfrom{\conf'}{\zstrat'}{\ostrat'}$ and since $\runfrom{\conf'}{\zstrat}{\ostrat} \models_\game \parity$ and since it is a valid run under the {\sc vass} semantics, we deduce that the values of the counters always remain positive. Consequently we have $\runfrom{\conf'}{\zstrat'}{\ostrat'} \models_\game \parity\wedge\always\nonneg$. We conclude that $[0,\energy]:\conf\models_\game\parity\wedge\always\nonneg$.

\subsection*{Proof of Lemma~\ref{lem:energy-vass}}

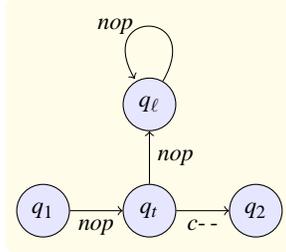
\begin{figure}[htbp]
\begin{center}
\begin{tikzpicture}[show background rectangle]
\node[name=dummy]{};
\node at (dummy)[state,name=q1]{$\state_1$};
\node at  ($(q1.east)+(7mm,0mm)$) [state,name=qt,anchor=west]{$\state_\transition$};
\node at  ($(qt.east)+(7mm,0mm)$) [state,name=q2,anchor=west]{$\state_2$};
\node at  ($(qt.north)+(0mm,7mm)$) [state,name=losing,anchor=south]{$\losingstate$};
\draw[->] (q1) -- node[below=0mm] {\footnotesize$\nop$} (qt);
\draw[->] (qt) -- node[below=0mm] {\footnotesize$\dec\cntr$} (q2);
\draw[->] (qt) -- node[textnode,right=0mm] {$\nop$} (losing);
\draw [->,in=120,out=60,loop ] (losing) edge node[textnode,left=1mm] {$\nop$} (losing);
\end{tikzpicture}
\caption{Translating a transition $\tuple{\state_1,\dec\cntr,\state_2}$
from an energy game to a single-sided {\sc vass} game.
Note that $\coloringof{\losingstate}$ is odd.}
\label{dec:transition:fig}
\end{center}
\end{figure}

Let $\game=\gametuple$ be an integer game. From it we build a single-sided integer game $\game'=\tuple{\stateset',\transitionset',\coloring'}$ as follows:
\begin{itemize}
\item $\stateset'=\stateset \uplus \set{\state_\transition \mid \transition \in \transitionset} \uplus \set{\losingstate}$ (where $\uplus$ denotes the disjoint union operator), with $\zstateset'=\zstateset \uplus \set{\state_\transition \mid \transition \in \transitionset} \uplus \set{\losingstate}$ and $\ostateset'=\ostateset$;
\item $\transitionset'$ is the smallest set of transitions such that, for each transition $\transition=\tuple{\state_1,\op,\state_2}$ in $\transitionset$, the following conditions are respected:
\begin{itemize}
\item $\tuple{\state_1,\nop,\state_\transition} \in \transitionset'$;
\item $\tuple{\state_\transition,\op,\state_2} \in \transitionset'$;
\item $\tuple{\state_\transition,\nop,\losingstate} \in \transitionset'$;
\item $\tuple{\losingstate,\nop,\losingstate} \in \transitionset'$;
\end{itemize}
\item $\coloring'$ is defined as follows:
\begin{itemize}
\item for all $\state \in \stateset$, $\coloring'(\state)=\coloring(\state)$;
\item for all $\transition \in \transitionset$, $\coloring'(\state_\transition)=0$;
\item $\coloring'(\losingstate)=1$.
\end{itemize}
\end{itemize}
By construction $\game'$ is single-sided. Also note that once the system enters the losing state $\losingstate$, Player 0 loses the game since the only possible infinite run from this state remains in $\losingstate$ and the color associated to this state is odd (it is equal to $1$). Figure \ref{dec:transition:fig} depicts the encoding of transitions of the form  $\tuple{\state_1,\dec\cntr,\state_2}$.

We will now prove that $\winsetpara{\game,\energy,0,\cntrs}{\parity\wedge\always\nonneg}=\winsetpara{\game',\vass,0,\cntrs}{\parity} \cap \set{\conf \mid \stateof{\conf} \in \stateset}$. First let $\conf \in \winsetpara{\game,\energy,0,\cntrs}{\parity\wedge\always\nonneg}$. This means that there exists $\conf' \in \instantiationsof{\conf}$ and  $\zstrat \in \energyzstratset$ such that $[0,\zstrat,\energy]:\conf\models_\game \parity\wedge\always\nonneg$. From $\zstrat$, we will build a winning strategy $\zstrat' \in \vasszstratset$ for player 0 in $\game'$. Let us first introduce some notation. To a path in $\game'$, $\path=\conf_0 \vmovesto{\transition_1} \conf_{\transition_1} \vmovesto{\transition'_1}
\conf_1 \vmovesto{\transition_2} \conf_{\transition_2} \vmovesto{\transition'_2} \conf_2
\cdots\conf_n$ with $\stateof{\conf_n} \in \stateset$, we associate the path $\betapath{\path}=\conf_0 \emovesto{\transition1}
\conf_1\emovesto{\transition2}
\cdots\conf_n$  in $\game$ (by construction of $\game'$ such a path exists). The strategy $\zstrat'$ is then defined as follows. For all paths $\path=\conf_0 \vmovesto{\transition_1} 
\conf_1 \vmovesto{\transition_2}  \conf_2
\cdots\conf_n$ in $\game'$:
\begin{itemize}
\item if $\stateof{\conf_n} \in \stateset$, then $\zstrat'(\path)=\tuple{\stateof{\conf_n},\nop,\state_\transition}$ with $\transition=\zstrat(\betapath{\path})$;
\item if $\stateof{\conf_n}=\state_\transition$ for some transition $\transition=\tuple{\state_1,\op,\state_2} \in \transitionset$, then if $\tuple{\state_\transition,\op,\state_2}$ is enabled in $\conf_n$, $\zstrat'(\path)=\tuple{\state_\transition,\op,\state_2}$, otherwise $\zstrat'=\tuple{\state_\transition,\nop,\losingstate}$;
\item if $\stateof{\conf_n} =\losingstate$, then $\zstrat'(\path)=\tuple{\losingstate,\nop,\losingstate}$.
\end{itemize} 
One can then easily verify using the definition of $\game'$ and of the strategy $\zstrat'$ that since $[0,\zstrat,\energy]:\conf\models_\game \parity\wedge\always\nonneg$, we have $[0,\zstrat',\vass]:\conf\models_\game \parity$ and hence that $\conf \in \winsetpara{\game',\vass,0,\cntrs}{\parity} \cap \set{ \conf \mid \stateof{\conf} \in \stateset}$.

The proof that if we take $\conf \in \winsetpara{\game',\vass,0,\cntrs}{\parity} \cap \set{ \conf \mid \stateof{\conf} \in \stateset}$ then $\conf$ belongs also to $\winsetpara{\game,\energy,0,\cntrs}{\parity \wedge \always \nonneg}$ is done similarly.

\subsection*{Proof of Lemma~\ref{lem:subconf}}

Let $\conf$ be a nonnegative partial configuration such that $\domof{\conf}=\cntrs' \subset \cntrs$.
Suppose that  $\cinstantiationsof\cntrs{\conf}\cap
\winsetpara{\game,\vass,0,\cntrs}{\parity}\neq\emptyset$, i.e.,
there is a $\conf_1\in\cinstantiationsof\cntrs{\conf}$
where $\conf_1\in\winsetpara{\game,\vass,0,\cntrs}{\parity}$.
Since $\conf_1\in\winsetpara{\game,\vass,0,\cntrs}{\parity}$
there is a 
$\conf_2\in\instantiationsof{\conf_1}$
with $[0,\vass]:\conf_2\models_\game\parity$.
Notice that $\conf_2\in\instantiationsof{\conf}$.
It follows that $\conf\in\winsetpara{\game,\vass,0,\cntrs'}{\parity}$.

Now, suppose that 
$\conf\in\winsetpara{\game,\vass,0,\cntrs'}{\parity}$.
By definition there is a $\conf_1\in\instantiationsof{\conf}$
such that  $[0,\vass]:\conf_1\models_\game\parity$.
Define $\conf_2$ by $\conf_2(\cntr):=\conf_1(\cntr)$
for all $\cntr\in\cntrs$ and $\conf_2(\cntr):=\undef$ for all $\cntr\notin\cntrs$.
Then $\conf_2\in\cinstantiationsof{\cntrs}{\conf}$ and
$\conf_2\in\winsetpara{\game,\vass,0,\cntrs}{\parity}$, hence $\cinstantiationsof\cntrs{\conf}\cap
\winsetpara{\game,\vass,0,\cntrs}{\parity}\neq\emptyset$.

\subsection*{Proof of Lemma~\ref{lem:coveredby}}

\begin{enumerate}
\item
Consider a partial nonnegative configuration $\hat{\conf} \in \confs$ where
$\cntr \in \cntrs - \domof{\hat{\conf}}$.

Since $\instantiationsof{\hat{\conf}} \cap
{\winsetpara{\game,\vass,0,\cntrset}{\parity}} \neq \emptyset$,
there exists a minimal finite number $\minpump(\hat{\conf})$
s.t. $\winset(\hat{\conf}) := \instantiationsof{\hat{\conf}[\cntr\assigned \minpump(\hat{\conf})]} \cap
{\winsetpara{\game,\vass,0,\cntrset}{\parity}} \neq \emptyset$.
\item
In particular, $\winset(\hat{\conf})$ is upward-closed w.r.t. the counters
in $\cntrset - \cntrs$ and
$\minofthis{\winset(\hat{\conf})}$ is finite.
Let $\minstart(\hat{\conf})$ be the maximal constant appearing in 
$\minofthis{\winset(\hat{\conf})}$.
Thus, an instantiation of $\hat{\conf}[\cntr\assigned \minpump(\hat{\conf})]$
where the counters in $\cntrset - \cntrs$ have values $\ge \minstart(\hat{\conf})$
is certainly winning for Player $0$, i.e., in
${\winsetpara{\game,\vass,0,\cntrset}{\parity}}$.
\item
The first condition of Def.~\ref{def:coveredby} is satisfied by the definition
of $\beta$.
Moreover, since $\conf \in\pconfset\cntrs$ and
$\instantiationsof{\conf} \cap {\winsetpara{\game,\vass,0,\cntrset}{\parity}} \neq
\emptyset$, for every $\cntr \in \cntrs$ we have 
$\instantiationsof{\conf[\cntr\assigned\undef]} \cap {\winsetpara{\game,\vass,0,\cntrset}{\parity}} \neq
\emptyset$.
Since $\beta$ are by definition the minimal nonnegative configurations (with a domain
which is exactly one element smaller than $\cntrs$) that have this property, 
there must exist some element $\hat{\conf} \in \confs$ s.t.
$\hat{\conf} \ordering \conf$. 
Therefore, also the second condition of Def.~\ref{def:coveredby} is 
satisfied and we get $\confs\coveredby\conf$. \qed
\end{enumerate}

\subsection*{Proof of Lemma~\ref{algorithm:termination:lemma}}

We assume the contrary and derive a contradiction.
If Algorithm~\ref{graph:algorithm} does not terminate then,  
in the graph of the game $\outgame$, it will build an infinite sequence of 
states $\state_0,\ldots,\state_k,\ldots$  such that,  for all $i,j \in \nat$, the following properties hold: $i < j$ implies 
\begin{description}
\item[(a)] $(\state_i,\state_j) \in \left(\outtransitionset\right)^*$, and,
\item[(b)] $\labelingof{\state_i} \neq \labelingof{\state_j}$, and,
\item[(c)] $\labelingof{\state_i} \not\sordering \labelingof{\state_j}$. 
\end{description}
The property (a) comes from the way we build the transition relation when
adding new state to the set to $\toexplore$ at Line 17 of the algorithm (and
from the fact that the {\sc vass} is finitely branching and hence so is the
graph of the game $\outgame$). The property (b) is deduced thanks to the test
at Line 12 that necessarily fails infinitely often, otherwise the algorithm
would terminate. 
The property (c) is obtained thanks to the test at Line 9 which must also fail
infinitely often if the algorithm does not terminate. 
Since the number of counters is fixed, the set $(\pconfset\cntrs,\ordering)$ 
is well-quasi-ordered by Dickson's Lemma.
Hence in the infinite sequence of states $\state_0,\ldots,\state_k,\ldots$
there must appear two states $\state_i$ and $\state_j$ with $i<j$ such that 
$\labelingof{\state_i} \ordering \labelingof{\state_j}$, 
which is a contradiction to the conjunction of (b) and (c). 
This allows us to conclude that the Algorithm~\ref{graph:algorithm} necessarily terminates.

\subsection*{Proof of Lemma~\ref{algorithm:pcorrectness:lemma}}

We show both directions of the equivalence.
\paragraph{Left to right implication.}
If $[0,\vass]:\conf\models_\game\parity$ then there exists a concrete
nonnegative configuration $\conf_0 \in \instantiationsof\conf$ 
with
$\conf_0 = \conf \union \conf'$ 
s.t.
$[0,\vass]:\conf_0 \models_\game\parity$, i.e.,
$\conf'$ assigns values to the counters in $\cntrset - \cntrs$.
Moreover, we have $\outconf=\tuple{\outstate,\cntrval_{\it out}}$ where 
$\labelingof\outstate = \conf$ and $\domof{\cntrval_{\it out}}=\emptyset$.

Using the winning strategy $\zstrat\in\vasszstratset$ of Player $0$ in
$\game$ from $\conf_0$, we will construct a winning strategy 
$\zstrat'\in\energyzstratset$ of Player $0$ from a concrete configuration
$\conf'_0 \in \instantiationsof\outconf$ in
$\outgame$, where $\conf'_0 = \tuple{\outstate,\cntrvalof{\conf'}}$.
We do this by maintaining a correspondence between nonnegative configurations
in both games and between the used sequences of transitions.
Let $\path=\conf_0\vmovesto{\transition_1}
\conf_1\vmovesto{\transition_2}\dots \conf_n$ a partial play in $\game$,
and
$\path' =\conf_0'\emovesto{\transition_1'}
\conf_1'\emovesto{\transition_2'}\dots \conf_n'$ a partial play in $\outgame$.

We will define $\zstrat'$ to ensure that either the following invariant holds for all $i \ge 0$ or
{\bf Condition 2} holds for some $\conf_n'$ and the invariant holds for all $i \le n$.
\begin{enumerate}
\item
$\labelingof{\transition_i'} = \transition_i$
\item
$\labelingof{\stateof{\conf_i'}} = \conf_i | \cntrs$
\item
$\cntrvalof{\conf_i'} = \conf_i | (\cntrset - \cntrs)$
\item
$\coloringof{\conf_i'} = \coloringof{\conf_i}$
\end{enumerate}
These conditions are satisfied for the initial states at $i=0$, since 
$\labelingof{\stateof{\conf_0'}} = 
\labelingof{\outstate} =
\conf =
\conf_0 | \cntrs$,
$\cntrvalof{\conf_0'} = \cntrvalof{\conf'}
= \conf_0 | (\cntrset - \cntrs)$
and
$\coloringof{\conf_0'} = 
\coloringof{\outstate} = 
\coloringof{\conf} =
\coloringof{\conf_0}$.

For the step we choose
$\zstrat'(\path') := t_{n+1}'$ s.t. 
$\labelingof{\transition_{n+1}'} = \transition_{n+1} = \zstrat(\path)$ which maintains the invariant.

It cannot happen that {\bf Condition 1} holds in $\path'$. All visited
nonnegative configurations $\conf_i$ in the winning play $\path$ are also winning for 
Player $0$. By Lemma~\ref{lem:coveredby} (item 3), we have
$\confs\coveredby\conf_i | \cntrs$
and thus $\confs \coveredby \conf_i | \cntrs =
\labelingof{\stateof{\conf_i'}}$ so that {\bf Condition 1} is false at $\conf_i'$.

Since $\game$ is a {\sc vass}-game, we have $\conf_i \ge 0$ for all $i\ge 0$.
Therefore 
$\labelingof{\stateof{\conf_i'}} = \conf_i | \cntrs \ge 0$
and
$\cntrvalof{\conf_i'} = \conf_i | (\cntrset - \cntrs) \ge 0$.
Thus the same transitions are possible in $\outgame$ as in $\game$.

In the case where {\bf Condition 2} eventually holds in $\outgame$, Player $0$ trivially wins
the game in $\outgame$.
Otherwise we have $\cntrvalof{\conf_i'} = \conf_i | (\cntrset - \cntrs) \ge 0$
for all $i\ge 0$ and thus the nonnegativity condition $\always\nonneg$ of
$\outgame$ is satisfied by $\path'$.

Finally, since the parity condition is satisfied by $\path$ and
$\coloringof{\conf_i} = \coloringof{\conf_i'}$, the parity condition is
also satisfied by $\path'$. Therefore $\zstrat'$ is winning for Player $0$ in
$\outgame$ from $\conf'_0 \in \instantiationsof\outconf$ and 
thus we obtain $[0,\energy]:\outconf\models_{\outgame}\parity\wedge\always\nonneg$ as
required.

\paragraph{Right to left implication.}
If $[0,\energy]:\outconf\models_{\outgame}\parity\wedge\always\nonneg$
then there exists a concrete
nonnegative configuration $\conf' \in \instantiationsof\outconf$ s.t.
$[0,\energy]:\conf'\models_{\outgame}\parity\wedge\always\nonneg$.
Due to the concreteness of $\conf'$ and the $\always\nonneg$ property,
we also have $[0,\vass]:\conf'\models_{\outgame}\parity$.
Thus Player $0$ has a winning strategy $\zstrat$ 
in the {\sc vass} parity game
on $\outgame$ from the concrete nonnegative configuration $\conf'$.

Using $\zstrat$, we will construct a winning strategy $\zstrat'$
for Player $0$ in the {\sc vass} parity game on $\game$ from some
nonnegative configuration $\conf_0 \in \instantiationsof\conf$.
Let $\conf_0 = \conf \union \conf''$, where $\conf''$ is some
yet to be constructed function assigning sufficiently high values to counters
in $\cntrset - \cntrs$.
We only prove the sufficient condition that a winning strategy 
$\zstrat'$ exists, but do not construct
a Turing machine that implements it. This is because $\zstrat'$
uses the numbers $\minpump(\hat{\conf})$ and $\minstart(\hat{\conf})$ from
Lemma~\ref{lem:coveredby} that are not computed here.

In order to construct $\zstrat'$ and $\conf''$, we need 
some definitions. 
Consider a sequence of transitions in $\outgame$ 
that leads from $\state$ to $\state'$ ending with {\bf Condition 2}
at line~\ref{win:line} in the algorithm.
We call this sequence a {\em pumping sequence}. Its effect is 
nonnegative on all counters in $\cntrs$ and strictly increasing in at least
one of them, although its effect may be negative on counters in 
$\cntrset - \cntrs$.
Due to the finiteness of $\outgame$ (by
Lemma~\ref{algorithm:termination:lemma}),
the number of different pumping sequences is bounded by some number $p$
and their maximal length is bonded by some number $l$.
For the given finite $\confs = \bigcup_{\cntrs' \subseteq \cntrs, |\cntrs'|=|\cntrs|-1}
\paretopara{\game,\vass,0,\cntrs'}{\parity}$ we use the constants from 
Lemma~\ref{lem:coveredby} to define the following finite upper bounds
$\minpump := \maxofthis{\{\minpump(\hat{\conf})\ |\ \hat{\conf} \in \confs\}}$
and
$\minstart := \maxofthis{\{\minstart(\hat{\conf})\ |\ \hat{\conf} \in \confs\}}$.

Now we define $\zstrat'$. The intuition is as follows.
Either the current nonnegative configuration is already known to be winning
for Player $0$ by induction hypothesis (if the current nonnegative configuration is
sufficiently large compared to nonnegative configurations in $\instantiationsof{\confs}$)
in which case he plays according to his known winning strategy from the
induction hypothesis.
Otherwise, for a given history $\path$ in $\game$,
Player $0$ plays like for a history $\path'$ in $\outgame$,
where $\path'$ is derived from $\path$ as follows.
For $\path'$ we first use a sequence of transitions in $\outgame$ 
whose labels (see line 13 of the algorithm) correspond to the sequence of transitions in
$\path$, but then we remove all subsequences from $\path'$ which are pumping
sequences in $\outgame$. Thus Player $0$ plays from nonnegative configurations in $\game$
that are possibly larger than the corresponding (labels of) nonnegative configurations in $\outgame$
on the counters in $\cntrs$. The other counters in $\cntrset - \cntrs$ might
differ between the games and will have to be chosen sufficiently high by the
initial $\conf''$ to stay positive during the game (see below).
We show that the history of the winning game in $\game$ will contain only finitely
many such pumping sequences, and thus finite initial values   
(encoded in $\conf''$) for the counters in $\cntrset - \cntrs$
will suffice to win the game.

Let $\path=\conf_0\movesto{\transition_1}
\conf_1\movesto{\transition_2}
\cdots\conf_n$ be a path in $\game$, where Player $0$ played according to
strategy $\zstrat'$.
Our strategy $\zstrat'$ will maintain the invariant that $\path$ induces a
sequence of states $\hat{\path} = \state_0, \state_1, \dots, \state_n$ in $\outgame$.
The sequence $\hat{\path}$ is almost like a path in $\outgame$ with
transitions whose label is the same as the transitions
in $\path$, except that it contains back-jumps to previously visited states 
whenever a pumping sequence is completed.

Let $\state_0 = \outstate$. For the step from
$\state_i$ to $\state_{i+1}$ there are two cases.
For a given transition $\transition_i$ in $\game$ appearing in $\path$ there is a unique transition
$\transition_i'$ in $\outgame$ with $\labelingof{\transition_i'} = \transition_i$. 
As an auxiliary construction we define the state $\state_{i+1}'$, which 
is characterized uniquely by $\labelingof{\state_{i+1}'} =
\transition_i'(\labelingof{\state_i})$.
If there is a $j \le i$ s.t.
the sequence from $\state_j$ to $\state_{i+1}'$ is a pumping sequence
and $\state_j$ is not part of a previously identified pumping sequence 
(the construction ensures that there can be at most one such $j$),
then let $\state_{i+1} := \state_j$, i.e., we jump back to the beginning of the 
pumping sequence.
Otherwise, if no pumping sequence is completed at $\state_{i+1}'$,
then let $\state_{i+1} = \state_{i+1}'$, so that we have 
$\state_i \movesto{\transition_i'} \state_{i+1}$.
From the sequence $\hat{\path}$ we obtain a genuine path 
$\path'$ in $\outgame$ by deleting all pumping sequences from
$\hat{\path}$.

In the case where $\conf_n$ belongs to Player $0$ we define
$\zstrat'(\path)$ by case distinction.
\begin{enumerate}
\item
We let $\zstrat'(\path) := \transition_i$ where $\labelingof{\transition_i} =
\zstrat(\path')$, except when the condition of the
following case 2 holds.
\item
By $\labelingof{\state_0} = \conf$ and $\conf_0 = \conf \union \conf''$
we have ${\labelingof{\state_0}} \ordering
\restrictedto{\cntrs}{\conf_0}$.
Since the effects of the sequences of transitions in $\path$ and $\hat{\path}$ are the same,
and pumping sequences have a nondecreasing effect on the counters in $\cntrs$,
we obtain ${\labelingof{\state_i}} \ordering
\restrictedto{\cntrs}{\conf_i}$ for all $i \ge 0$.

Since $\zstrat$ is winning in $\outgame$ we have
$\confs\coveredby\labelingof{\state_i}$
and thus $\confs\coveredby \restrictedto{\cntrs}{\conf_i}$.
By Def.~\ref{def:coveredby},
there exists some $\hat{\conf} \in \confs$ and counter $\cntr \notin
\domof{\hat{\conf}}$ s.t.
$\hat{\conf}\ordering\restrictedto{\cntrs}{\conf_i}[\cntr\assigned\undef]$.

{\bf Condition for case 2:} 
If $\conf_i(\cntr) \ge \minpump(\hat{\conf})$ and 
$\conf_i(\cntr') \ge \minstart(\hat{\conf})$ for every counter $\cntr' \in
\cntrset - \cntrs$ then,
by Lemma~\ref{lem:coveredby} (items 1 and 2) and
monotonicity (Lemma~\ref{monotonicity:lemma}), 
Player $0$ has a winning strategy $\zstrat''$ from $\conf_i$.
In this case $\zstrat'$ henceforth follows this winning strategy $\zstrat''$.
\end{enumerate}

Now we show that $\zstrat'$ is winning for Player $0$ in $\game$ from the
initial nonnegative configuration $\conf_0 = \conf \union \conf''$ for some
sufficiently large but finite $\conf''$.
We distinguish two cases, depending on whether case 2 above is reached or not.

{\bf If Case 2 is reached:}
Consider the case where condition 2 above holds at some reached game 
nonnegative configuration $\conf_n$.
Every pumping sequence $\alpha$ has nondecreasing effect on all counters in
$\cntrs$ and strictly increases at least some counter $\cntr_\alpha \in
\cntrs$.
Thus if $\hat{\path}$ contains the pumping sequence $\alpha$ at least $\minpump$ times,
then $\conf_n(\cntr_\alpha) - \labelingof{\state_n}(\cntr_\alpha) \ge \minpump$ and
in particular $\conf_n(\cntr_\alpha) \ge \minpump$.
If additionally, $\conf_n$ is sufficiently large on the counters outside $C$, 
i.e., $\conf_n(\cntr') \ge \minstart$ for every counter $\cntr' \in
\cntrset - \cntrs$, then 
case 2 above applies and the winning strategy $\zstrat''$ takes over.

The path $\path$ (resp. $\hat{\path}$) can contain at most 
$\minpump * p$ pumping sequences of a combined length that is bounded 
by $\minpump * p * l$ before the first condition $\conf_n(\cntr_\alpha) \ge
\minpump$ becomes true for some pumping sequence $\alpha$. In this case 
it is sufficient for $\zstrat''$ to win if the values in the counters in 
$\cntrset - \cntrs$ are $\ge \minstart$ at nonnegative configuration $\conf_n$.
How large does a counter $\cntr' \in \cntrset - \cntrs$ need to be at the
(part of the) initial nonnegative configuration $\conf''$ in order to satisfy this
additional condition later at $\conf_n$?
Since $\zstrat$ is winning in the {\sc vass} game from $\conf'$ in $\outgame$,
an initial value $\conf'(\cntr')$ is sufficient to keep the counter $\cntr'$ above $0$
in the game on $\outgame$. Thus an  initial value of
$\conf'(\cntr')+\minstart$ is sufficient to keep the counter
$\cntr'$ above $\minstart$ in the game on $\outgame$.
Moreover, the game played according to $\zstrat'$ in $\game$ contains the same
transitions (modulo the labeling $\labelingof{\dots}$) 
as the game played according to $\zstrat$ on $\outgame$, except
for the $\le v*p*l$ extra transitions in pumping sequences.
Since a single transition can decrease a counter by at most one, 
an initial counter value of 
$\conf''(\cntr') = \conf'(\cntr')+ \minstart + \minpump * p * l$ 
is sufficient in order to have $\cntr' \ge \minstart$ whenever case 2 applies
and then $\zstrat''$ (and thus $\zstrat'$) is winning for Player $0$.
The counters in $\cntrs$ are always large enough by construction, since 
${\labelingof{\state_i}} \ordering \restrictedto{\cntrs}{\conf_i}$ for 
all $n \ge i \ge 0$.
The parity objective is satisfied by $\zstrat'$, since it is satisfied by
$\zstrat''$ on the infinite suffix of the game.

{\bf If Case 2 is not reached:}
Otherwise, if case 2 is not reached, then the {\sc vass} game on $\game$
played according to $\zstrat'$ is like
the {\sc vass} game on $\outgame$ played according to $\zstrat$, 
except for the finitely many interludes of pumping sequences, of
which there are at most $p*\minpump$ 
(with a combined length $\le \minpump * p * l$).
Since $\zstrat$ is winning the {\sc vass} game on $\outgame$ from $\conf'$,
this keeps the counters nonnegative.
At most $\minpump * p * l$ extra transitions happen in $\game$
(in the pumping sequences) and a single transition can decrement a 
counter by at most one.
Thus it is sufficient for staying nonnegative in $\game$ if $\conf''(\cntr') \ge
\conf'(\cntr')+\minpump * p * l$ for all $\cntr' \in \cntrset - \cntrs$. 
The counters in $\cntr \in \cntrs$ trivially stay nonnegative, since 
${\labelingof{\state_i}} \ordering \restrictedto{\cntrs}{\conf_i}$ for all $i \ge 0$.
The parity objective is satisfied, since the colors of the nonnegative configurations
$\conf_i$ and $\state_i$ in $\path$ and $\hat{\path}$ coincide,
the colors of an infinite suffix of $\hat{\path}$ coincide with the colors
of an infinite suffix of $\path'$ and $\path'$ satisfies the parity
objective as $\zstrat$ is winning in $\outgame$.

{\bf Combination of the cases.}
While $\zstrat'$ might not be able to enforce either of the two cases
described above, one of them will certainly hold in any play.
We define the (part of the) initial nonnegative configuration $\conf''$ to be
sufficiently high to win in either case, by taking the maximum of the 
requirements for the cases.

We let $\conf''(\cntr') := \conf'(\cntr)+ \minstart + \minpump * p * l$ 
for all $\cntr' \in \cntrset - \cntrs$
and obtain that $\zstrat'$ is a winning strategy for Player $0$ in the parity
game on $\game$ from the
initial nonnegative configuration $\conf_0 = \conf \union \conf'' \in
\instantiationsof{\conf}$.
Thus $[0,\vass]:\conf\models_\game\parity$, as required. \qed

\subsection*{Proof of Theorem~\ref{thm:weaksim}}

Given a labeled finite-state system $\fstuple$ and a labeled {\sc vass}
$\vasstuple$ with initial states $\fsstate_0$ and
$\tuple{\state_0,\cntrval}$, respectively,
we construct a single-sided integer game $\game=\tuple{\stateset_0 \uplus \stateset_1,\transitionset',\coloring}$
with initial configuration $\conf=\tuple{\tuple{\fsstate_0,\state_0, 1}, \cntrval}$ s.t.
$\tuple{\state_0,\cntrval}$ weakly simulates $\fsstate_0$ if and only if
$[0,\vass]:\conf\models_\game\parity$. Then decidability follows from
Theorem~\ref{single:sided:vass:parity:theorem}.

Let $\stateset_1 = \{\tuple{\fsstate,\state, 1}\ |\ \fsstate \in \fsstates,
\state \in \stateset\} \cup \{{\it win}_0\}$ and
$\stateset_0 = \{\tuple{\fsstate,\state, 0} \ |\ \fsstate \in \fsstates,
\state \in \stateset\} \cup
\{\tuple{\fsstate,\state^a, 0} \ |\ \fsstate \in \fsstates,
\state \in \stateset, a \in \alphabet\} \cup \{{\it lose}_0\}$.
Let $\coloringof{\stateset_1} =2$ and $\coloringof{\stateset_0}=1$, i.e., Player
$0$ wins the parity game iff states belonging to Player $1$ are visited
infinitely often. 

Now we define $\transitionset'$.
For every finite-state system transition $\fsstate \movesto{a} \fsstate'$ and every $\state \in \stateset$,
we add a transition
$\tuple{\tuple{\fsstate,\state, 1}, \nop, \tuple{\fsstate',\state^a, 0}}$.
Here the state $\state^a$ encodes the choice of the symbol $a$ by Player $1$,
which restricts the future moves of Player $0$.
For every {\sc vass} transition
$\transition=\tuple{\state_1,\op,\state_2}\in\transitionset$ with label
$\labelingof{\transition} = \tau$ and every $\fsstate \in \fsstates, a\in \alphabet$ we add a transition
$\tuple{\tuple{\fsstate,\state_1^a, 0}, \op, \tuple{\fsstate,\state_2^a, 0}}$.
This encodes the first arbitrarily long sequence of $\tau$-moves in the Player
$0$ response of the form $\tau^* a \tau^*$. 
For every {\sc vass} transition
$\transition=\tuple{\state_1,\op,\state_2}\in\transitionset$ with label
$\labelingof{\transition} = a \neq \tau$ and $\fsstate \in \fsstates$ we add a transition
$\tuple{\tuple{\fsstate,\state_1^a, 0}, \op, \tuple{\fsstate,\state_2, 0}}$.
This encodes the $a$-step in in the Player
$0$ response of the form $\tau^* a \tau^*$. 
Moreover, we add transitions
$\tuple{\tuple{\fsstate,\state^\tau, 0}, \nop, \tuple{\fsstate,\state, 0}}$ for
all $\fsstate \in \fsstates,\state \in \stateset$ (since a $\tau$-move in the
weak simulation game does
not strictly require a response step).
For every {\sc vass} transition
$\transition=\tuple{\state_1,\op,\state_2}\in\transitionset$ with label
$\labelingof{\transition} = \tau$ and $\fsstate \in \fsstates$ we add a transition
$\tuple{\tuple{\fsstate,\state_1, 0}, \op, \tuple{\fsstate,\state_2, 0}}$.
This encodes the second arbitrarily long sequence of $\tau$-moves in the Player
$0$ response of the form $\tau^* a \tau^*$.
Finally, for all $\fsstate \in \fsstates, \state \in \stateset$ we add
transitions
$\tuple{\tuple{\fsstate,\state, 0}, \nop, \tuple{\fsstate,\state, 1}}$.
Here Player $0$ switches the control back to Player $1$. He cannot win by
delaying this switch indefinitely, because the color of the states in
$\stateset_0$ is odd.

The following transitions encode the property of the simulation game
that a player loses if he gets stuck.
For every state in $\state \in \stateset_1$ with no outgoing transitions we
add a transition $\tuple{\state, \nop, {\it win}_0}$. 
In particular this creates a loop at state ${\it win}_0$.
Since the color of ${\it win}_0$ is even, this state is winning for
Player $0$.
For every state in $\state \in \stateset_0$ with no outgoing transitions we
add a transition $\tuple{\state, \nop, {\it lose}_0}$.
In particular this creates a loop at state ${\it lose}_0$.
Since the color of ${\it lose}_0$ is odd, this state is losing for
Player $0$.

This construction yields a single-sided integer game, since all transitions
from states in $\stateset_1$ have operation $\nop$.

A round of the weak simulation game is encoded by the moves of the players
between successive visits to a state in $\stateset_1$. A winning strategy for
Player $0$ in the weak simulation game directly induces a winning strategy for
Player $0$ in the parity game $\game$, since the highest color that is
infinitely often visited is $2$, and thus $[0,\vass]:\conf\models_\game\parity$.
Conversely, $[0,\vass]:\conf\models_\game\parity$ implies a winning strategy
for Player $0$ in the parity game on $\game$ which ensures that
color $2$ is seen infinitely often. Therefore, states in $\stateset_1$ are visited
infinitely often. Thus, either infinitely many rounds of the weak simulation
game are simulated or state ${\it win}_0$ is reached in $\game$ and Player $1$
gets stuck in the weak simulation game. In either case, Player $0$ wins the
weak simulation game and $\tuple{\state_0,\cntrval}$ weakly simulates
$\fsstate_0$.
\qed

\subsection*{Semantics of $\posmucalcul$}

The syntax of the positive $\mu$-calculus $\posmucalcul$ is given
by the following grammar:
$\formula ::=  \state ~\mid~ \var ~\mid~ \formula \wedge \formula ~\mid~
\formula \vee \formula ~\mid~ \eventually \formula ~\mid ~ \always \formula
~\mid~ \least \var.\formula ~\mid~ \greatest \var.\formula$
where $\state \in \stateset$ and $\var$ belongs to a countable set of
variables $\varset$. 

Free and bound occurrences of variables are defined as
usual. We assume that no variable has both bound and free occurrences in some
$\phi$, and that no two fixpoint subterms bind the same variable (this can
always be ensured by renaming a bound variable). A formula is closed if it has
no free variables.
Without restriction, we do not use any negation in our syntax. Negation can be
pushed inward by the usual dualities of fixpoints, and
the negation of an atomic proposition referring to a control-state can be
expressed by a disjunction of propositions referring to all the other
control-states. 

We now give the interpretation over the {\sc vass}
$\unlabvasstuple$ of a formula of $\posmucalcul$ according to an environment
$\env : \varset \rightarrow 2^\confset$ which associates to each variable
a subset of concrete configurations. Given $\env$, a formula $\formula \in
\posmucalcul$ represents a subset of concrete configurations,
denoted by $\interpretationsof{\formula}_{\env}$ and defined inductively as follows.
$$
\begin{array}{lcl}
\interpretationsof{\state}_{\env} & = & \set{\conf \in \confset \mid \stateof{\conf}=\state}\\
\interpretationsof{\var}_{\env} &=& \env(\var) \\
\interpretationsof{\formula \wedge \formulabis}_{\env} & = & \interpretationsof{\formula}_{\env} \cap \interpretationsof{\formulabis}_{\env} \\
\interpretationsof{\formula \vee \formulabis}_{\env} & = & \interpretationsof{\formula}_{\env} \cup \interpretationsof{\formulabis}_{\env} \\
\interpretationsof{\eventually \formula}_{\env} & = & \set{\conf \in \confset \mid \exists \conf' \in \interpretationsof{\formula}_{\env} \mbox{~s.t.~} \conf \vmovesto{} \conf' } \\
\interpretationsof{\always \formula}_{\env} & = & \set{\conf \in \confset \mid \forall \conf' \in \confset,  \conf \vmovesto{} \conf' \mbox{ implies } \conf' \in \interpretationsof{\formula}_{\env}} \\
\interpretationsof{\least \var.\formula}_{\env} & = &\bigcap \set{\confset' \subseteq \confset \mid \interpretationsof{\formula}_{\env[\var \assigned \confset']} \subseteq \confset'}\\
\interpretationsof{\greatest \var.\formula}_{\env} & = &\bigcup \set{\confset' \subseteq \confset \mid \confset' \subseteq  \interpretationsof{\formula}_{\env[\var \assigned \confset']} }\\
\end{array}
$$
where the notation $\env[\var \assigned \confset']$ is used to define an
environment equal
to $\env$ on every variable except on $\var$ where it returns $\confset'$. We
recall that $(2^\confset,\subseteq)$ is a complete lattice and that, for every
$\formula \in \posmucalcul$ and every environment $\env$, the function $\func :
2^\confset \mapsto 2^\confset$, which associates to $\confset' \subseteq
\confset$ the set $\func(\confset')=\interpretationsof{\formula}_{\env[\var
    \assigned \confset']}$, is monotonic. Hence, by the Knaster-Tarski
Theorem, the set  $\interpretationsof{\least \var.\formula}_{\env}$
(resp. $\interpretationsof{\greatest \var.\formula}_{\env}$) is the least
fixpoint (resp. greatest fixpoint) of $\func$, and it is well-defined. Finally we denote by $\interpretationsof{\formula}$ the subset of configurations $\interpretationsof{\formula}_{\env_0}$ where $\env_0$ is the
environment which assigns the empty set to each variable.

\subsection*{Proof of Lemma~\ref{lem:mucalcul:to:parity}}

We consider a {\sc vass} $\avass=\unlabvasstuple$ and $\formula$ a  formula in $\posmucalcul$. We will use in this proof the set of subformulae of $\formula$, denoted by $\subformulae{\formula}$. For formulae in $\posmucalcul$ we assume that no variable is bounded by the same fixpoint. Hence given a formula $\formula$ and a bounded variable $\var \in \varset$, we can determine uniquely the subformula of $\formula$ that bounds the variable $\var$; such a formula will be denoted by $\formula_\var$. We also denote by $\freevarsof{\formula}$ the set of free variables in $\formula$.  The integer game $\game(\avass,\formula)=\tuple{\stateset',\transitionset',\coloring}$ is built as follows:
\begin{itemize}
\item $\stateset'=\stateset \times \subformulae{\formula}$
\item The transition relation $\transitionset'$ is the smallest set respecting the following conditions for all the formulae $\formulabis \in \subformulae{\formula}$:
\begin{itemize}
\item If $\formulabis=\state$ with $\state \in \stateset$, then $\tuple{\tuple{\state',\formulabis},\nop,\tuple{\state',\formulabis}}$ belongs to $\transitionset'$ for all states $\state'$ in $\stateset$;
\item If $\formulabis=\var$ with $\var \in \varset$ and $\var \notin \freevarsof{\formula}$, then $\tuple{\tuple{\state,\formulabis},\nop,\tuple{\state,\formula_\var}}$ belongs to $\transitionset'$ for all states $\state$ in $\stateset$;
\item If $\formulabis=\var$ with $\var \in \varset$ and $\var \in \freevarsof{\formula}$, then $\tuple{\tuple{\state,\formulabis},\nop,\tuple{\state,\formulabis}}$ belongs to $\transitionset'$ for all states $\state$ in $\stateset$;
\item If $\formulabis=\formulabis' \wedge \formulabis''$ or $\formulabis=\formulabis' \vee \formulabis''$ then  $\tuple{\tuple{\state,\formulabis},\nop,\tuple{\state,\formulabis'}}$ and  $\tuple{\tuple{\state,\formulabis},\nop,\tuple{\state,\formulabis''}}$ belong to $\transitionset'$ for all states $\state$ in $\stateset$;
\item If $\formulabis=\eventually \formulabis'$ or $\formulabis=\always \formulabis'$   then for all states $\state \in \stateset$ and for all transitions $\tuple{\state,\op,\state'} \in \transitionset$, we have $\tuple{\tuple{\state,\formulabis},\op,\tuple{\state',\formulabis'}}$ in $\transitionset'$;
\item If $\formulabis=\least \var.\formulabis'$ or $\formulabis= \greatest \var.\formulabis'$, then  $\tuple{\tuple{\state,\formulabis},\nop,\tuple{\state,\formulabis'}}$ for all states $\state \in \stateset$.
\end{itemize}
\item A state $\tuple{\state,\formulabis}$ belongs to $\zstateset'$ if and only if:
\begin{itemize}
\item $\formulabis=\state'$ with $\state' \in \stateset$, or,
\item $\formulabis=\var$ with $\var \in \varset$, or,
\item $\formulabis=\formulabis' \vee \formulabis''$, or,
\item $\formulabis=\eventually \formulabis'$, or,
\item $\formulabis=\least \var.\formulabis''$, or,
\item $\formulabis=\greatest \var.\formulabis''$.
\end{itemize}
\item A state $\tuple{\state,\formulabis}$ belongs to $\ostateset'$ if and only if:
\begin{itemize}
\item $\formulabis=\formulabis' \wedge \formulabis''$, or,
\item $\formulabis=\always \formulabis'$.
\end{itemize}
\item The coloring function $\coloring$ is then defined as follows:
\begin{itemize}
\item for all $\state,\state' \in \stateset$, if $\state'=\state$ then $\coloring{\tuple{\state,\state'}}=0$  and if $\state' \neq \state$  then $\coloring{\tuple{\state,\state'}}=1$;
\item for all $\state \in \stateset$ and all $\var \in \freevarsof{\formula}$, $\coloring{\tuple{\state,\var}}=1$;
\item for all $\state \in \stateset$, for all subformulae $\formulabis \in \subformulae{\formula}$ if $\formulabis \neq \least \var.\formulabis''$ and $\formulabis \neq \greatest \var.\formulabis''$ and $\formulabis \neq \state'$ with $\state' \in \stateset$ and $\formulabis \neq \var$ with $\var \in \freevarsof{\formula}$ , then $\coloring{\tuple{\state,\formulabis}}=0$;
\item for all $\state \in \stateset$, for all subformulae $\formulabis \in \subformulae{\formula}$ such that $\formulabis \neq \least \var.\formulabis''$,  $\coloring{\tuple{\state,\formulabis}}=m$ where $m$ is the smallest odd number greater or equal to the alternation depth of $\formulabis$;
\item for all $\state \in \stateset$, for all subformulae $\formulabis \in \subformulae{\formula}$ such that $\formulabis \neq \least \var.\formulabis''$,  $\coloring{\tuple{\state,\formulabis}}=m$ where $m$ is the smallest even number greater or equal to the alternation depth of $\formulabis$;
\end{itemize}
\end{itemize}
Before providing the main property of the game $\game(\avass,\formula)$, we introduce a new winning condition which will be useful in the sequel of the proof. This winning condition uses an environment  $\env : \varset \rightarrow 2^\confset$ and is given by the formula $\parity \vee \bigvee_{\var \in \freevarsof{\formula}} \eventually (\var \wedge \env(\var)) $ where  $\var \wedge \env(\var)$ holds in the configurations of the form $\tuple{\tuple{\state,\var},\cntrval}$ such that $\tuple{\state,\cntrval}\in \env(\var)$. It states that a run is winning if it respects the parity condition or if at some point it encounters a configuration of the form $\tuple{\tuple{\state,\var},\cntrval}$ with $\var \in \freevarsof{\formula}$ and $\tuple{\state,\cntrval} \in \env(\var)$. We denote by $\condition(\formula,\env)$ the formula $\bigvee_{\var \in \freevarsof{\formula}} \eventually (\var \wedge \env(\var)) $.

We will now prove the following property: for all formulae $\formula$ in $\posmucalcul$, for all concrete configurations $\conf=\tuple{\state,\cntrval}$ of $\avass$ and all environments $\env : \varset \rightarrow 2^\confset$, we have $\conf \in \interpretationsof{\formula}_{\env}$ iff $[0,\vass]:\tuple{\tuple{\state,\formula},\cntrval} \models_{\game(\avass,\formula)}\parity \vee \condition(\formula,\env) $, i.e., iff $\tuple{\tuple{\state,\formula},\cntrval} \in \winsetpara{\game(\avass,\formula),\vass,0,\cntrset}{\parity \vee \condition(\formula,\env)}$.

We reason by induction on the length of $\formula$. For the base case with $\formula=\state$ with $\state \in \stateset$ or $\formula=\var$ with $\var \in \varset$ the property trivially holds. We then proceed with the induction reasoning. It is  easy to prove that the property holds for formulae of the form $\formula' \wedge \formula''$ or $\formula' \vee \formula''$ if the property holds for $\formula'$ and $\formula''$ and the same for formulae of the form $\eventually \formula'$ and $\always \formula'$. We consider now a formula $\formula$ of the form $\least \var.\formulabis$ and assume that the property holds for the formula $\formulabis$. Let $\func : 2^\confset \mapsto 2^\confset$ be the function which associates to any subset of configurations $\confset'$ the set $\func(\confset')=\interpretationsof{\formulabis}_{\env[\var \assigned \confset']}$. By induction hypothesis we have $\tuple{\state,\cntrval} \in \func(\confset')$ iff $\tuple{\tuple{\state,\formulabis},\cntrval} \in \winsetpara{\game(\avass,\formulabis),\vass,0,\cntrset}{\parity \vee \condition(\formulabis,\env[\var \assigned \confset'])}$. We denote by $\least\func$ the least fixpoint of $\func$. We want to prove that $\tuple{\state,\cntrval} \in \least\func$ iff $\tuple{\tuple{\state,\least \var.\formulabis},\cntrval} \in \winsetpara{\game(\avass,\least \var.\formulabis),\vass,0,\cntrset}{\parity \vee \condition(\least \var.\formulabis,\env)}$. We define the following set of configurations $\confset_\least=\set{\tuple{\state,\cntrval} \in \confset \mid \tuple{\tuple{\state,\least \var.\formulabis},\cntrval} \in \winsetpara{\game(\avass,\least \var.\formulabis),\vass,0,\cntrset}{\parity \vee \condition(\least \var.\formulabis, \env)}}$. So finally what we want to prove is that $\least\func=\confset_\least$.
\begin{itemize}
\item We begin by proving that $\least\func \subseteq \confset_\least$. By definition $\least\func = \bigcap \set{\confset' \subseteq \confset \mid \func(\confset') \subseteq \confset'}$. Hence it is enough to prove that $\func(\confset_\least) \subseteq \confset_\least$. Let $\tuple{\state,\cntrval} \in \func(\confset_\least)$. This means that $\tuple{\tuple{\state,\formulabis},\cntrval} \in \winsetpara{\game(\avass,\formulabis),\vass,0,\cntrset}{\parity \vee \condition(\formulabis,\env[\var \assigned \confset_\least])}$ by definition of $\func$. We want to prove that $\tuple{\tuple{\state,\least \var.\formulabis},\cntrval} \in \winsetpara{\game(\avass,\formulabis),\vass,0,\cntrset}{\parity\vee \condition(\least \var.\formulabis,\env)}$. First note that the configuration $\tuple{\tuple{\state,\least \var.\formulabis},\cntrval}$  belongs to Player 0, and from this configuration, Player 0 has a unique choice which is to go to the state $\tuple{\tuple{\state,\formulabis},\cntrval}$. Then from $\tuple{\tuple{\state,\formulabis},\cntrval}$, if Player $0$ plays as in the game $\game(\avass,\formulabis)$ where it has a winning strategy, there are two options:
\begin{enumerate}
\item a control-state of the form $\tuple{\state',\var}$ is never encountered and in that case Player 0 wins because it was winning in $\game(\avass,\formulabis)$ and the run performed is the same;
\item a control-state of the form $\tuple{\state',\var}$ is encountered, but in that case, Player 0 is necessarily in a configuration $\tuple{\tuple{\state',\var},\cntrval}$ with $\tuple{\state',\cntrval} \in \env[\var \assigned \confset_\least](\var)$ (by definition of the winning condition in $\game(\avass,\formulabis)$), ie with $\tuple{\state',\cntrval} \in \confset_\least$. But this means that from this configuration, Player 0 has a winning strategy for the game $\game(\avass,\formula)$.
\end{enumerate}
Hence we have shown that $\tuple{\state,\cntrval} \in \confset_\least$ and consequently  $\func(\confset_\least) \subseteq \confset_\least$. This allows us to deduce that $\least\func \subseteq \confset_\least$.
\item We will now prove that $\confset_\least \subseteq \least\func$. For this we will prove that for all $\confset' \subseteq \confset$ such that $\func(\confset')=\confset'$, we have $\confset_\least \subseteq \confset'$. This will in fact imply that $\confset_\least \subseteq \least\func$, since $\least\func$ is the least fixpoint of the function $\func$. Let $\confset' \subseteq \confset$ such that $\func(\confset')=\confset'$ and let $\tuple{\state,\cntrval} \in \confset_\least$. We reason \emph{by contradiction} and assume that $\tuple{\state,\cntrval} \notin \confset'$. Since $\tuple{\state,\cntrval} \in \confset_\least$, this means that Player 0 has a winning strategy to win in the game $\game(\avass,\least \var.\formulabis))$ from the configuration $\tuple{\tuple{\state,\least \var.\formulabis},\cntrval}$ with the objective $\parity \vee \condition(\least \var.\formulabis, \env)$. Since $\tuple{\state,\cntrval} \notin \confset'=\func(\confset')$, this means that there is no winning strategy for Player 0 in the game $\game(\avass,\formulabis)$ from configuration $\tuple{\tuple{\state,\formulabis},\cntrval}$ with the objective $\parity \vee \condition(\formulabis,\env[\var \assigned \confset'])$. Since  Player 0 has a winning strategy to win in the game $\game(\avass,\least \var.\formulabis)$, we can adapt this strategy to the game $\game(\avass,\formulabis)$ (by restricting it to the path possible in this game and beginning one step later). But since this strategy is not winning in the game $\game(\avass,\formulabis)$ with the objective $\parity \vee \condition(\formulabis,\env[\var \assigned \confset'])$, it means that there is a path $\path_0$ in $\game(\avass,\formulabis)$ that respects this strategy and this path necessarily terminates in a state of the form $\tuple{\tuple{\state_1,\var},\cntrval_1}$ with $\tuple{\state_1,\cntrval_1} \notin \env[\var \assigned \confset'](\var)$, i.e., with $\tuple{\state_1,\cntrval_1} \notin \confset'$ (otherwise this strategy which is winning in $\game(\avass,\least \var.\formulabis)$ would also be winning in $\game(\avass,\formulabis)$). On the other hand, in $\game(\avass,\least \var.\formulabis)$, $\tuple{\tuple{\state_1,\var},\cntrval_1}$ has a unique successor which is  $\tuple{\tuple{\state_1,\least \var.\formulabis},\cntrval_1}$ and from which Player 0 has a winning strategy since we have followed a winning strategy in the game $\game(\avass,\least \var.\formulabis)$ that has lead us to that configuration. Hence we have $\tuple{\state_1,\cntrval_1} \notin \confset'$ and $\tuple{\state_1,\cntrval_1} \in \confset_\least$. So from $\tuple{\state_1,\cntrval_1}$ we can perform a similar reasoning following the winning strategy in $\game(\avass,\least \var.\formulabis)$ to reach a configuration $\tuple{\tuple{\state_2,\var},\cntrval_2}$ such that $\tuple{\state_2,\cntrval_2} \notin \confset'$ and $\tuple{\state_2,\cntrval_2} \in \confset_\least$. Finally, by performing the same reasoning we succeed in building an infinite play in $\game(\avass,\least \var.\formulabis)$ which follows a winning strategy and such that the sequence of the visited configurations is of the form:
$$
\tuple{\tuple{\state,\least \var.\formulabis},\cntrval} \ldots \tuple{\tuple{\state_1,\least \var.\formulabis},\cntrval_1} \ldots \tuple{\tuple{\state_2,\least \var.\formulabis},\cntrval_2} \ldots 
$$

Note that for all $i \geq 1$, $\coloring(\tuple{\tuple{\state_i,\least \var.\formulabis},\cntrval_i})$ is the maximal priority in the game $\game(\avass,\least \var.\formula)$ and it is odd by definition of the game. This means that the path we obtain following a winning strategy for Player 0 is losing, which is a contradiction. Hence we have $\tuple{\state,\cntrval} \in \confset'$. From this we deduce that $\confset_\least \subseteq \least\func$.
\end{itemize}
If we consider a formula  $\formula$  of the form $\greatest \var.\formulabis$, a reasoning similar to the previous one can be performed in order to show that the property holds.

Thanks to the previous proof, for all formulae $\formula$ in $\posmucalcul$, for all concrete configurations $\conf=\tuple{\state,\cntrval}$ of $\avass$, we have $\conf \in \interpretationsof{\formula}_{\env_0}$ iff  $\tuple{\tuple{\state,\formula},\cntrval} \in \winsetpara{\game(\avass,\formula)),\vass,0,\cntrset}{\parity \vee \condition(\formula,\env_0)}$ where $\env_0$ is the environment which assigns to each variable the empty set. This means that for all formulae $\formula$ in $\posmucalcul$, for all concrete configurations $\conf=\tuple{\state,\cntrval}$ of $\avass$, we have $\conf \in \interpretationsof{\formula}_{\env_0}$ iff  $\tuple{\tuple{\state,\formula},\cntrval} \in \winsetpara{\game(\avass,\formula)),\vass,0,\cntrset}{\parity}$ because $\condition(\formula,\env_0)$ is equivalent to the formula which is always false. By denoting $\conf'=\tuple{\tuple{\state,\formula},\cntrval}$, we have hence that $[0,\vass]:\conf' \models_{\game(\avass,\formula)}\parity$ if and only if $\avass,\conf \models \formula$.

\subsection*{Proof of Lemma~\ref{lem:mucalcul:single:to:parity}}
Let $\avass=\unlabvasstuple$ be a single-sided {\sc vass} and $\formula \in \guardmucalcul$. Strictly speaking the construction of the game  $\game(\avass,\formula)$ proposed in the proof of Lemma \ref{lem:mucalcul:to:parity} does not build a single-sided game. However we can adapt this construction in order to build an equivalent single-sided game. In this manner we adapt the construction to the case of $\formula \in \guardmucalcul$ by changing the rules for the outgoing transitions for states in the game of the form $\tuple{\state,\ostateset \wedge \always \formulabis}$. To achieve this we build a game $\game'(\avass,\formula)=\tuple{\stateset',\transitionset',\coloring}$ the same way as $\game(\avass,\formula)$ except that we perform the following change  in the definition of transition relation for states of the form $\tuple{\state,\ostateset \wedge \always \formulabis}$:
\begin{itemize}
\item If  $\formulabis=\ostateset \wedge \always \formulabis'$   then for all states $\state \in \ostateset$, for all transitions $\tuple{\state,\nop,\state'} \in \transitionset$, we have $\tuple{\tuple{\state,\formulabis},\nop,\tuple{\state',\formulabis'}}$ in $\transitionset'$, and, for all states $\state \in \zstateset$, we have $\tuple{\tuple{\state,\formulabis},\nop,\tuple{\state,\formulabis}}$ in $\transitionset'$.
\end{itemize}
Then the states of the form $\tuple{\state,\ostateset \wedge \always \formulabis}$ will belong to Player 1 and the coloring of such states will be defined as follows:
\begin{itemize}
\item for all $\state \in \stateset$, for all subformulae $\formulabis \in \subformulae{\formula}$, if $\formulabis=\stateset \wedge \always \formulabis'$ then if $\state \in \ostateset$, $\coloring(\tuple{\state,\formulabis})=0$ else $\coloring(\tuple{\state,\formulabis})=1$.
\end{itemize}
Apart from these changes the definition of the game $\game'(\avass,\formula)$ is equivalent to the one of $\game(\avass,\formula)$. By construction, since $\avass$ is single-sided and by definition of $\guardmucalcul$, we have that such an integer game $\game'(\avass,\formula)$ is single-sided. Furthermore, for any concrete configuration $\conf=\tuple{\tuple{\state,\formulabis},\cntrval}$, one can easily show that $[0,\vass]:\conf \models_{\game(\avass,\formula)}\parity$ iff $[0,\vass]:\conf \models_{\game'(\avass,\formula)}\parity$.

\subsection*{Proof of Theorem~\ref{thm:mucalcul:decidable}}
Let $\avass=\unlabvasstuple$ be a single-sided {\sc vass},  $\formula$ be
closed formula of $\guardmucalcul$ and $\conf_0$ be an initial configuration
of $\unlabvasstuple$. Using Lemma~\ref{lem:mucalcul:to:parity} 
and Lemma~\ref{lem:mucalcul:single:to:parity}, we have that $[0,\vass]:\conf'_0
\models_{\game'(\avass,\formula)}\parity$ if and only if $\avass,\conf_0
\models \formula$ where $\game'(\avass,\formula)$ is a single-sided integer
game. Hence, thanks to Corollary \ref{corollary:vass:decidable}, we can deduce
that the model-checking problem of $\guardmucalcul$ over single-sided {\sc vass} is decidable. Furthermore, by using the result of these two lemmas we have  that $\tuple{\state,\cntrval} \in \interpretationsof{\formula}_{\env_0}$ iff $\tuple{\tuple{\state,\formula},\cntrval} \in \winsetpara{\game'(\avass,\formula),\vass,0,\cntrs}{\parity}$. Hence by Corollary \ref{cor:vass-energy}, we deduce that $\interpretationsof{\formula}_{\env_0}$ is upward-closed and by Theorem \ref{single:sided:vass:parity:theorem} that we can compute its set of minimal elements which is equal to $\set{\tuple{\state,\cntrval} \mid 
\tuple{\tuple{\state,\formula},\cntrval} \in  \paretopara{\game'(\avass,\formula),\vass,0,\cntrset}{\parity}}$.

\end{document}